\documentclass[aps,prx,twocolumn,superscriptaddress,nofootinbib,a4paper]{revtex4-2}

\usepackage{CJK}
\usepackage[pdftex]{graphicx}
\usepackage{amsmath}
\usepackage{verbatim}
\usepackage{xcolor}
\usepackage{bm}
\usepackage{txfonts}
\usepackage{siunitx}
\usepackage[colorlinks=true,allcolors=blue]{hyperref} 
\usepackage{multirow}

\begin{document}

\title{Long-range optomechanical interactions in SiN membrane arrays}\thanks{This work was published in \href{https://doi.org/10.1103/PhysRevX.15.011014}{Phys.\ Rev.\ X \textbf{15}, 011014 (2025).}}

\author{Xiong Yao}
\affiliation{Department of Physics, Fudan University, Shanghai 200438, P.R.\ China}
\affiliation{Faculty of Physics, School of Science, Westlake University, Hangzhou 310030, P.R.\ China}
\affiliation{Kavli Institute of Nanoscience, Department of Quantum Nanoscience, Delft University of Technology, Lorentzweg 1, 2628CJ Delft, The Netherlands}

\author{Matthijs H.\ J.\ de Jong}\thanks{Present address:\ Department of Applied Physics, Aalto University, FI-00076 Aalto, Finland}
\affiliation{Kavli Institute of Nanoscience, Department of Quantum Nanoscience, Delft University of Technology, Lorentzweg 1, 2628CJ Delft, The Netherlands}
\affiliation{Department of Precision and Microsystems Engineering, Delft University of Technology, Mekelweg 2, 2628CD Delft, The Netherlands}

\author{Jie Li}
\affiliation{School of Physics, Zhejiang University, Hangzhou 310027, P.R.\ China}

\author{Simon Gr\"{o}blacher}
\email[]{s.groeblacher@tudelft.nl}
\affiliation{Kavli Institute of Nanoscience, Department of Quantum Nanoscience, Delft University of Technology, Lorentzweg 1, 2628CJ Delft, The Netherlands}

\begin{abstract}
Optomechanical systems using a membrane-in-the-middle configuration can exhibit a long-range type of interaction similar to how atoms show collective motion in an optical potential. Photons bounce back and forth inside a high-finesse Fabry-P\'{e}rot cavity and mediate the interaction between multiple membranes over a significant distance compared to the wavelength. Recently, it has been demonstrated that off-resonant coupling between light and the inter-membrane cavity can lead to coherent mechanical noise cancellation. On-resonance coupling of light with both the Fabry-P\'{e}rot and inter-membrane cavities, predicted to enhance the single photon optomechanical coupling, have to date not been experimentally demonstrated, however. In our experiment, a double-membrane system inside a Fabry-P\'{e}rot cavity resonantly enhances the cavity field, resulting in a stronger optomechanical coupling strength from the increased radiation pressure. The resonance condition is first identified by analyzing the slope of the dispersion relation. Then, the optomechanical coupling is determined at various chip positions over one wavelength range. The optimum coupling conditions are obtained and enhancement is demonstrated for double membrane arrays with three different reflectivites, reaching nearly four-fold enhancement for the collective motion of $R=65\%$ double membranes. The cavity losses at the optimum coupling are also characterized and the potential of reaching the single-photon strong coupling regime is discussed.
\end{abstract}

\maketitle

\section*{Introduction}

Membrane-in-the-middle optomechanics first received attention due to its ability to independently engineer the optical cavities and mechanical resonators~\cite{thompson2008strong}. Many exciting experiments have been realized over the years using single membranes, including optomechanical ground state cooling~\cite{Peterson2016PRL}, sensing~\cite{reinhardt2016ultralow,halg2021membrane}, mode squeezing~\cite{nielsen2017multimode, huang2024room}, entanglement~\cite{thomas2021entanglement}, and an optomechanical memory~\cite{MadsPRL2024}. Extending the system to multiple membranes inside a high-finesse Fabry-P\'{e}rot (FP) cavity enables many additional opportunities to test new physics, using long-range optomechanical interactions~\cite{Xuereb2012,xuereb2013collectively,xuereb2014reconfigurable}. In such an experiment, the light field mediates mechanical motion between multiple modes, leading to effects such as hybridization~\cite{Shkarin2014} and synchronization~\cite{Bemani2017PRA, Sheng2020} of mechanical motion, topological~\cite{Xu2016} and cavity-mediated heat transport~\cite{yang2020phonon}, coherent state transfer~\cite{Weaver2017}, and mechanical noise cancellation ~\cite{de2022coherent, Marzioni2023}. 

One of the most exciting prospects of such a multi-membrane system is the ability to realize single-photon strong optomechanical coupling, where the single-photon optomechanical coupling strength $g_\mathrm{0}$ is larger than both optical loss $\kappa$ and mechanical dissipation $\gamma_\mathrm{M}$~\cite{Xuereb2012,xuereb2013collectively}. In this regime, the nonlinear nature of the optomechanical coupling $\hbar g_0 \hat{a}^{\dagger} \hat{a} \left(\hat{b}^{\dagger}+\hat{b}\right)$ becomes dominant and the typical linearized form of the Hamiltonian breaks down~\cite{Aspelmeyer2014,bowen2015quantum}. Here $\hat{a}$, $\hat{a}^{\dagger}$ and $\hat{b}$, $\hat{b}^{\dagger}$ are photonic and phononic annihilation and creation operators, respectively. In this regime, phenomena such as the optomechanical photon blockade~\cite{rabl2011photon} and the generation of non-Gaussian mechanical states~\cite{nunnenkamp2011single} will become observable. Furthermore, strong single-photon coupling could lead to enhanced optomechanical squeezing~\cite{safavi2013squeezed}, which is beneficial for quantum sensing~\cite{nielsen2017multimode,aggarwal2020room}. One of the most promising routes to this regime for membrane-in-the-middle systems is to enhance the cavity field between multiple high-reflectivity membranes, where multi-membrane systems collectively interact with a single optical mode~\cite{Xuereb2012,xuereb2013collectively,Li2016,Newsom2020}. The strongly localized light field only couples to the breathing mode of a stack of identical membranes~\cite{Xuereb2012,Li2016,Newsom2020}. 

Despite this exciting prospect, no clear observation of this effect has been made to date. Typically, the optomechanical coupling rate $g_\mathrm{0}$ of a membrane-in-the-middle system can only reach up to the order of a few Kilohertz due to the large optical cavity mode volume~\cite{pfeifer2022achievements, Saarinen:23}, making it extremely challenging to reach $g_\mathrm{0}\!>\kappa$. Achieving $g_\mathrm{0}\!>\gamma_\mathrm{M}$ on the other hand has become relatively straightforward due to advanced mechanical engineering techniques such as high-stress material~\cite{norte2016mechanical,beccari2022strained}, soft-clamping~\cite{tsaturyan2017ultracoherent}, and phononic shields~\cite{enzian2023phononically}. Up until now, experimental efforts to demonstrate coupling enhancement of multiple membranes inside a FP cavity have only shown an increase of the linear optomechanical coupling $G$~\cite{Gaertner2018,Piergentili2018} measured via the slope of the dispersion curve. This way, $G$ is larger due to the multiple membranes acting as a single scatterer with increased response to the field. However, the single-photon coupling rate $g_\mathrm{0}$ is not increased in this case. When operating on resonance with the inter-membrane cavity, on the other hand, the light field is focused between the membranes and strongly couples to their collective motion, this long-range interaction leads to an enhanced collective coupling strength $g_\mathrm{0}$. One of the major experimental challenges in observing this effect lies in stabilizing these high-finesse cavities, which has only been solved recently~\cite{de2022coherent}.

In this work, we experimentally explore long-range type of optomechanical interactions which allow to enhance the collective coupling strength $g_\mathrm{0}$ when the resonance conditions of the outer high-finesse and inter-membrane cavities are met. We first introduce our integrated double-membrane (DM) system and its collective motion and verify the resonance conditions with the flat dispersion relations. Subsequently, the cavity is locked on resonance and mechanical spectra at various optical coupling powers are measured to extract the optomechanical coupling strength. This allows us to calibrate the coupling enhancement and cavity losses at each chip position. Additionally, by measuring devices with three different reflectivities, we benchmark our experimental performance against theoretical predictions~\cite{xuereb2013collectively,Newsom2020}. Finally, we also identify discrepancies between assumptions in the existing theory work~\cite{Xuereb2012,xuereb2013collectively,Li2016,Newsom2020} and our experiments in long-range collective motion, which is due to fabrication imperfections resulting in non-identical membranes and discussed in detail in the Appendix.

\begin{figure}
	\includegraphics[width = 0.48\textwidth]{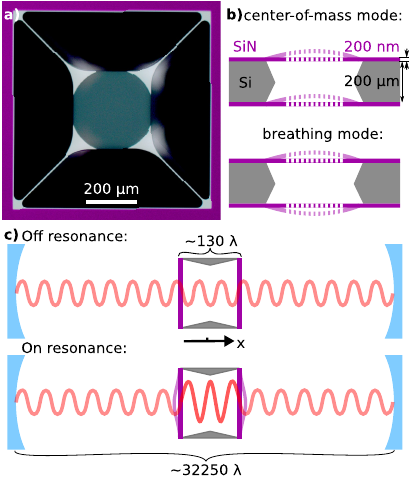}
	\caption{\textbf{a)} Microscope image of a SiN double-membrane trampoline device. The membrane on the backside is visible as a white shadow. The lateral offset between membranes on the front- and backside of the chip is less than \SI{35}{\micro\meter}, which is much smaller than the extent of the PhC pad and does not cause significant optical losses, as the cavity beam waist is only \SI{33}{\micro\meter}. \textbf{b)} Side-view schematic of the identical membrane array device and collective mechanical oscillation in orthogonal basis, where the top panel shows the mechanical center-of-mass mode (oscillates in-phase), while the bottom panel shows the mechanical breathing mode (oscillates with opposite phase). \textbf{c)} Schematic of the optical field off- (top) and on-resonance (bottom) with the inter-membrane cavity. The light field increases inside the inter-membrane cavity compared to the off-resonance case, yielding a higher radiation pressure across both membranes and resulting in an enhanced optomechanical coupling strength.}
	\label{fig: Schematics}
\end{figure}

\section*{Results}
\subsection*{Integrated optomechanical array inside Fabry-P\'{e}rot Cavity}
\label{sec: devices}

Our devices are patterned into \SI{200}{nm} high-stress silicon nitride (SiN) films on both sides of a \SI{200}{\micro m} silicon (Si) substrate used as a spacer. Potassium hydroxide (KOH) etching of the substrate around the devices gives rise to an inter-membrane FP cavity, with a free spectral range (FSR) about \SI{6}{nm}, or \SI{750}{GHz} at the operating wavelength of \SI{1550}{nm}. The mechanical trampoline resonator designs we use here have been optimized in previous works~\cite{norte2016mechanical,Gaertner2018}, allowing us to control the optical reflectivity ($R$) anywhere from the intrinsic film value (approx.\ \SI{35}{\%}) to \SI{99.8}{\%} through design choices of a photonic crystal (PhC), while simultaneously reaching a mechanical quality factor $Q_{\mathrm{M}} \approx \! 10^6$. For this particular set of experiments, we fabricate devices with $R$ of \SI{35}{\%}, \SI{50}{\%}, and \SI{65}{\%} at \SI{1550}{nm}, respectively. The two trampolines in each device have nearly identical mechanical frequencies, with the fundamental mode (out-of-plane) between $\SI{111}{}$ and $\SI{114}{kHz}$~\cite{norte2016mechanical}. We attribute the residual spread to fabrication imperfections and slightly different PhC parameters. The top- and side-views of our double optomechanical array are shown in Fig.~\ref{fig: Schematics}a and b, respectively. More details about the devices are provided in the Appendix.

One of the key features of our device design is the single-substrate configuration, which allows for a highly uniform gap between the two membranes, avoiding alignment difficulties present in other experiments~\cite{Piergentili2018,Gaertner2018,Sheng2020}. The chip is positioned near the center of our \SI{49.6}{\milli\meter} long free-space high-finesse FP cavity~\cite{de2022coherent}, with a FSR of about \SI{24.2}{pm} (equivalent to \SI{3.02}{GHz} at \SI{1550}{nm}). The empty FP cavity has a total linewidth of $\kappa_\mathrm{empty}/2\pi \approx$~\SI{120}{\kilo\hertz}, which corresponds to a finesse of $\sim$25,000. A piezoelectric crystal is placed below the membrane chip, which allows for precise positioning of the chip along the optical axis of the FP cavity ($x$-direction) over multiple wavelengths (\SI{6}{\micro m} range) (see Fig.~\ref{fig: Schematics}c). For all practical purposes, our system remains a membrane-in-the-middle and not a membrane-close-to-the-end-mirror system, even at maximum displacement, which may otherwise restrict the light to the region between one membrane and the cavity mirror, rather than between the two membranes~\cite{stambaugh2015membrane,Newsom2020}

\begin{figure*}[t]
	\includegraphics[width = 0.9\textwidth]{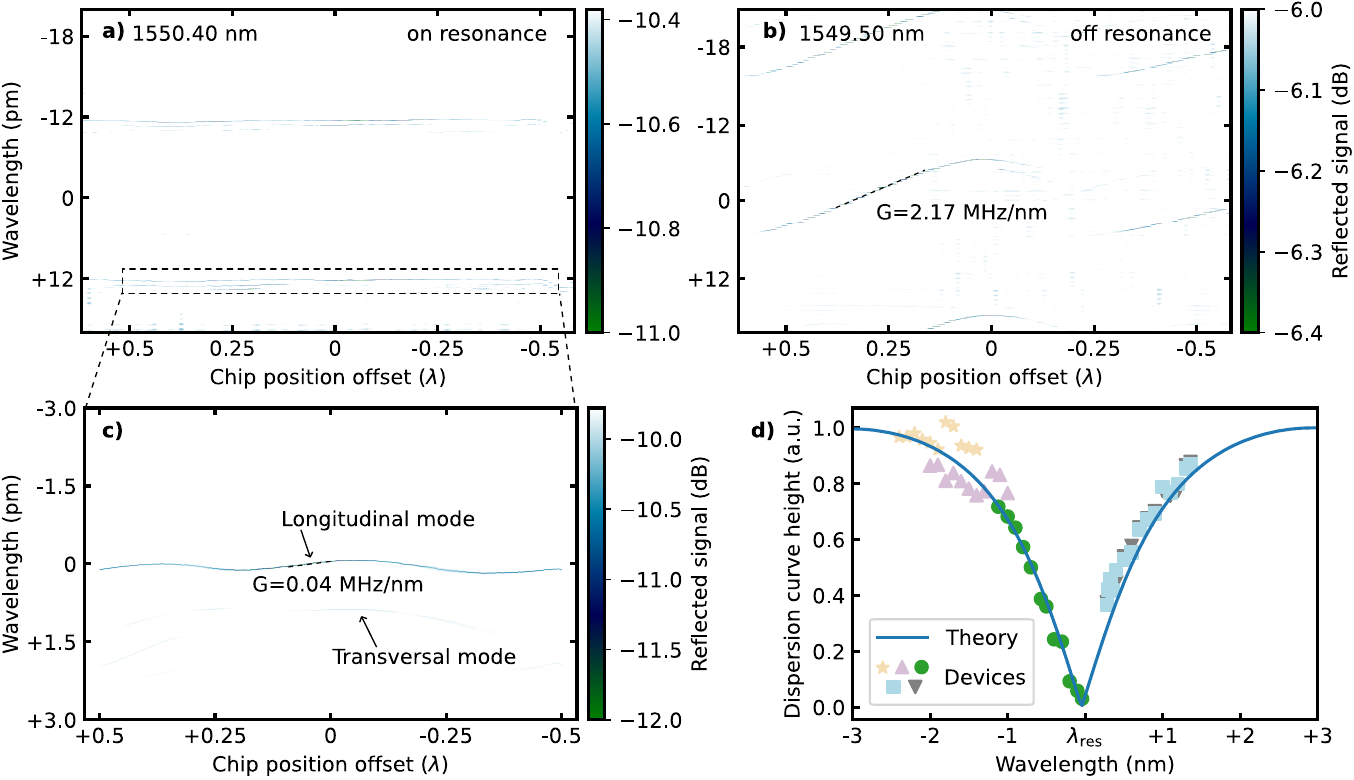}
	\caption{Characterization of the $R = 35\%$ double membrane device. \textbf{a)} Dispersion curve close to the resonance of the inter-membrane cavity. \textbf{b)} Dispersion curve off-resonance. The maximum linear coupling strength is \SI{2.17}{MHz/nm}. \textbf{c)} Zoom-in of the dispersion curve from a), showing that the dispersion tends to flatten (\SI{0.04}{MHz/nm}) when $\lambda$ is close to the resonance of the inter-membrane cavity. The additional first transversal mode of the cavity is due to small alignment imperfections. a) and b) are measured with a broader wavelength scan of approximately \SI{42}{pm}. c) is a finer scan, to accurately identify the inter-membrane resonance condition ($\lambda_\mathbf{res}$). \textbf{d)} Normalized dispersion curve height vs.\ input wavelength. The blue solid line is the numerical simulation of a fixed membrane spacing based on \cite{Li2016}. The different data points represent the normalized height of the dispersion curve measured for different membranes when scanning the wavelength of the laser. One device (green, round dots) is on-resonance within the laser operating wavelength range. $\lambda_\mathrm{res}$ is \SI{1550.41}{nm}, shown in a) and c).}
	\label{fig: Dispersion_curve}
\end{figure*}

\subsection*{Inter-membrane cavity resonance}
\label{sec: dispersion}

In order to observe the enhanced single photon optomechanical coupling of the two-membrane system, the operating wavelength of our laser has to simultaneously match both the resonance conditions of the main cavity as well as the inter-membrane cavity. This way, the field strength is redistributed. We identify matching both resonance conditions by analyzing the slope of the dispersion curves, i.e.\ the maximum linear optomechanical coupling $G = \mathrm{max}(|\partial \omega_\mathrm{c} /\partial x|)$~\cite{thompson2008strong}. The $G$ vanishes when we match both resonance conditions, unlike for the case of a single membrane (SM) ~\cite{Li2016,Gaertner2018,Riedinger2018}. This behavior is mainly due to the $\partial \omega_\mathrm{c} /\partial x$ of each membrane having opposite signs when the resonant condition is met. Consequently, the net cavity frequency shift cancels out in dispersion curves.

Since our membranes are less reflective than the free-space FP cavity mirrors ($>$99.9\%), we predominantly find resonances of the main cavity. The dispersion curves of both SM and DMs are periodic with $\lambda/2$~\cite{thompson2008strong,jayich2008dispersive,Gaertner2018, Wei2019PRA}. Figs.~\ref{fig: Dispersion_curve}a-c show the on- and off-resonance dispersion curves of the $R=0.35$ DMs, respectively. When off-resonance, the dispersion curve exhibits a large variation of cavity resonance frequency $\omega_\mathrm{c}$ as a function of the membrane position (height of dispersion curve in Fig.~\ref{fig: Dispersion_curve}d). The largest slope yields a coupling $G=$~\SI{2.17}{MHz/nm} near \SI{1549.50}{nm}. Conversely, on-resonance we observe a flat dispersion curve with a maximum coupling strength of only $G=$~\SI{0.04}{MHz/nm} near \SI{1550.45}{nm}. This prevents us to directly extract $g_0$ for the collective motion where the enhancement occurs. Fig.~\ref{fig: Dispersion_curve}c also shows the first transversal cavity mode, predominantly due to imperfect mode-matching between the incident laser beam and cavity, as well as small imperfections in alignment of the DMs stack with the main cavity. In general, the alignment of the DMs devices within the cavity is technically challenging, and greatly exacerbated if the membranes are highly reflective.

The dispersion curve can be modeled by the transfer matrix method, including two dielectric slabs between two mirrors~\cite{Li2016}. With the same parameters, we obtain dispersion curves in exactly the same manner as we do in the experiment (see details in the Appendix). We refer to the difference between minimum and maximum cavity frequency as the dispersion curve height~\cite{footnote1}. The blue curve in Fig.~\ref{fig: Dispersion_curve}d is our simulated result for $R$ of 35\%, which reaches zero when both cavities are on resonance. The predicted dip in dispersion curve height matches the inter-membrane cavity resonance that can be observed from a direct optical characterization of the membrane array~\cite{Nair2017,Gaertner2018}, and the width of this feature is determined by the finesse $\mathcal{F} \simeq 3$ of the inter-membrane cavity.

Our \SI{1}{\nano\meter} laser wavelength tuning range is much less than the \SI{6}{\nano\meter} inter-membrane cavity FSR, meaning we cannot see a full oscillation of the dispersion curve height in a single device. However, due to very small variations in the thickness across the chip on the order of $<1$~\si{\micro\meter}, different devices have distinct inter-membrane cavity resonance frequencies. For one device (green dots in Fig.~\ref{fig: Dispersion_curve}d) the inter-membrane cavity resonance condition falls within the tuning range of our laser. The other two devices, matching the laser wavelength tuning range for the DMs with higher reflectivities $R$ (\SI{50}{\%}, \SI{65}{\%}), are shown in the Appendix.

\begin{figure}[t]
	\includegraphics[width = 0.48\textwidth]{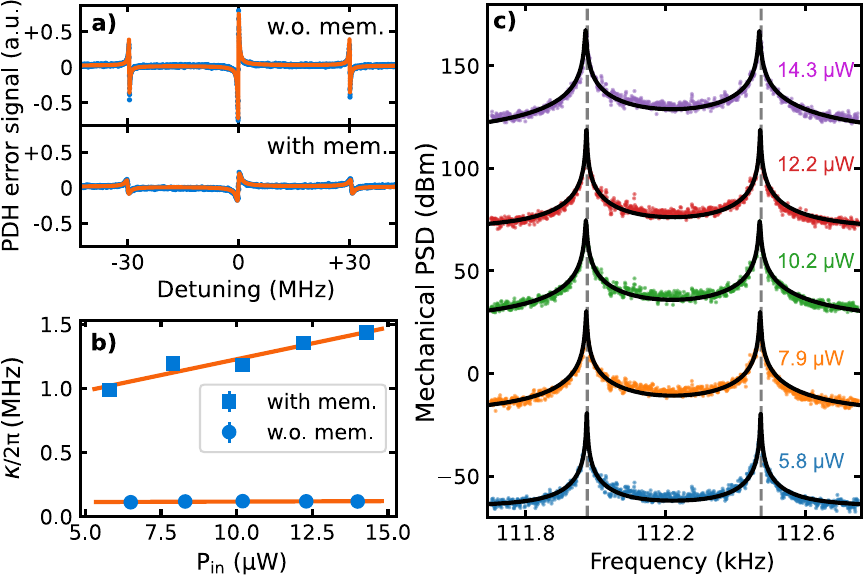}
	\caption{Optical and mechanical characterization of DM with $R=\SI{35}{\%}$. \textbf{a)} PDH error signal without (top) and with a membrane (bottom), both using the same y-axis scale. Blue dotted lines are raw data and the orange lines are a fit. \textbf{b)} Dependence of the cavity linewidth on optical input power. The lines (orange) show a linear regression. Error bars represent standard deviation obtained from the fit. \textbf{c)} Mechanical spectra showing the two fundamental modes of the membranes, all characterized at the chip position $0.25\lambda$ (cf.\ Fig.~\ref{fig: Dispersion_curve}c). The spectra are equally vertically shifted for visualization. The two gray vertical dashed lines indicate the intrinsic fundamental modes of the trampoline membranes.}
	\label{fig: Optomechanical_spectra}
\end{figure}

\subsection*{Optomechanical coupling characterization}
\label{sec: g0}

In order to obtain the single-photon optomechanical coupling rate $g_\mathrm{0}$, we measure the mechanical spectra with different input laser powers, from which we can directly extract the linearized optomechanical coupling $g = \sqrt{n_\mathrm{c}} g_\mathrm{0}$, where the mechanical frequency shifts due to the optical spring effect~\cite{Aspelmeyer2014}. It is important to keep the optical power relatively low to avoid optical bistability in the membrane-in-the-middle (MIM) system~\cite{enzian2023phononically}. The cavity photon number $n_\mathrm{c}$ can then be calculated by independently measuring the incident power, cavity mode-matching, cavity linewidth $\kappa$ and detuning $\Delta$~\cite{Aspelmeyer2014, de2022coherent}. The mechanical spectra are obtained through a homodyne detection scheme, combined with a Pound-Drever-Hall (PDH) technique locking the laser to the cavity resonance. The mechanical responses are fitted with a theoretical description based on a standard optomechanical Hamiltonian with two mechanical modes, as described in detail in~\cite{de2022coherent} and the Appendix.

Fig.~\ref{fig: Optomechanical_spectra} shows an exemplary set of measurements required to extract $g_\mathrm{0}$. We first measure the cavity linewidth $\kappa$ from fitting the PDH error signal, Fig.~\ref{fig: Optomechanical_spectra}a. Subtracting the external decay rate (empty FP cavity linewidth), $\kappa_\mathrm{empty}$, from the total $\kappa/2\pi$ $\simeq$ 560 kHz, we obtain internal losses due to the membrane, $\kappa_\mathrm{int}/2\pi$ $\simeq$ 440 kHz. Unlike $\kappa_\mathrm{empty}$, we observe that $\kappa_\mathrm{int}$ is power dependent, cf.\ Fig.~\ref{fig: Optomechanical_spectra}b. This may be attributed to an increase in diffraction~\cite{Monticone2017,peralle2024quasibound} due to a deformation of the DMs when increasing the optical power, which can be further exaggerated due to the alignment imperfection. We therefore measure $\kappa$ for each power and use it to compensate for the power-dependence. We obtain the $g_\mathrm{0}/2\pi$ for each of the two membranes $1.58 \pm 0.01 \mathrm{Hz}$ and $1.62 \pm 0.01 \mathrm{Hz}$, respectively, which is comparable to~\cite{de2022coherent}.

\begin{figure}
\includegraphics[width = 0.44\textwidth]{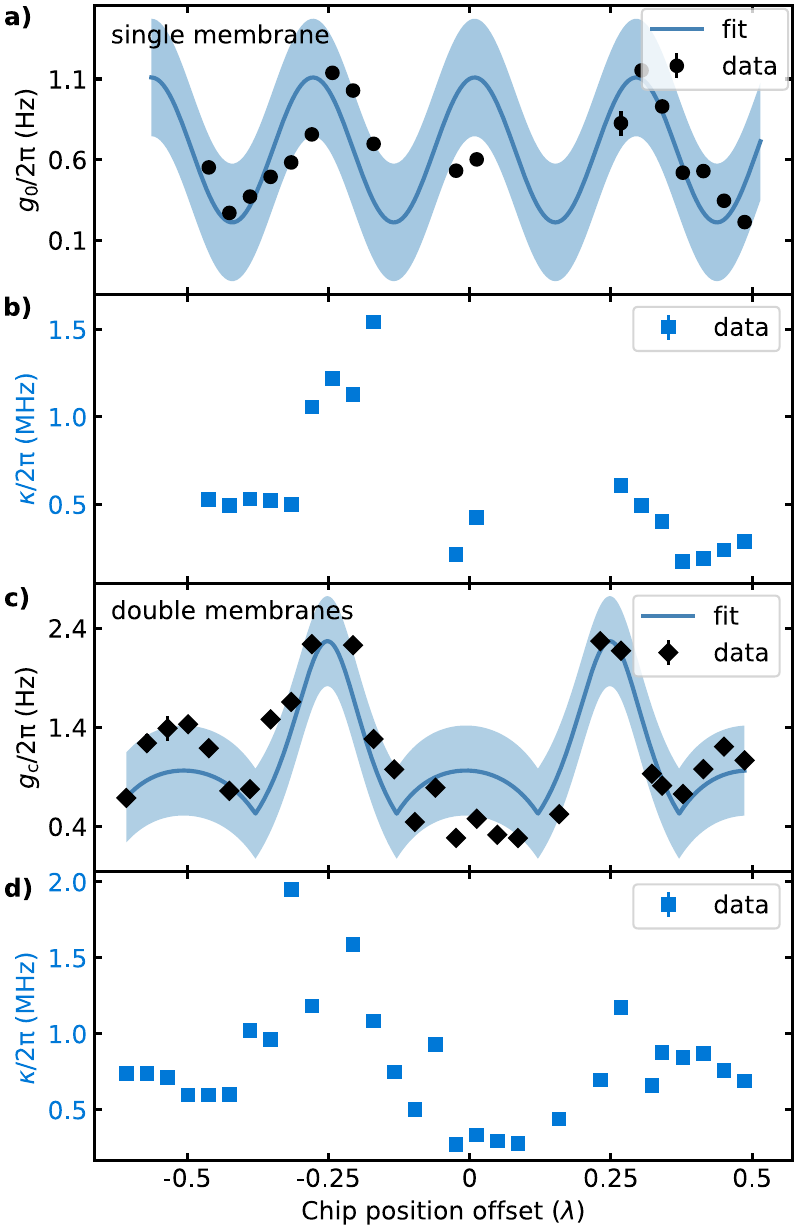}
\caption{Optomechanical coupling strength $g_\mathrm{0}$ ($g_\mathrm{c}$ for DMs) and cavity loss $\kappa$ as a function of the chip position. \textbf{a)} and \textbf{b)} are for SM and \textbf{c)} and \textbf{d)} are for DM, respectively. $g_\mathrm{0}$ is fitted by $|\sin(\theta/2)|^2$ for the SM and by Eq.~\eqref{eq:g0_norm} for the DMs. The blue shaded area in a) and c) indicates the fitting uncertainty of $g_\mathrm{0}$. Error bars represent standard deviation obtained from the fit.}
\label{fig:g0andkappa}
\end{figure}

The difference in light intensity on either side of the membrane gives rise to the radiation pressure that leads to the optomechanical coupling. The coupling strength for each individual membrane $g_\mathrm{0,j}$ in the array can be evaluated by~\cite{Xuereb2012,Li2016,Newsom2020}
\begin{equation}
    g_\mathrm{0, j} \propto x_\mathrm{zpf} \frac{\omega}{L}|I_\mathrm{R}-I_\mathrm{L}|, \quad j = 1, 2,
    \label{eq:intensity_diff}
\end{equation}
where 1 and 2 represent either membrane in the array. It is important to note that due to long-range interactions nature~\cite{Xuereb2012,xuereb2013collectively}, even in the weak coupling regime, each peak does not correspond to an individual membrane resonance. Therefore, we evaluate the collective coupling strength, $g_\mathrm{c}$, of the collective motion, which is predicted in theory~\cite{Newsom2020} and given by
\begin{equation}
    g_\mathrm{c} = \sqrt{g_\mathrm{0, 1}^2 + g_\mathrm{0, 2}^2}.
    \label{eq:intensity_diff}
\end{equation}
By incrementally moving the whole chip over the range of one wavelength (see Fig.~\ref{fig:g0andkappa}), we can find the position where the coupling is maximal through the field enhancement~\cite{Xuereb2012}. Comparing the case between a DM and a SM, for the latter the $g_\mathrm{0}$ of $R=35\%$ follows a $|\sin(\theta/2)|^2$ function, i.e., quarter-wavelength periodicity, and the maximum is found to be $g_\mathrm{0}/2\pi \approx 1.15\pm 0.03 \mathrm{Hz}$ (see Appendix). Conversely, we see the coupling rate $g_\mathrm{c}/2\pi$ of the DMs vary significantly, ranging from a minimum $0.28 \pm 0.02 \mathrm{Hz}$ to a maximum $2.27 \pm 0.07 \mathrm{Hz}$, exhibiting half-wavelength periodicity, consistent with theoretical predictions for the normalized coupling rate $g_\mathrm{c, norm}$~\cite{Newsom2020}
\begin{equation}
    g_\mathrm{c, norm} = \left| \frac{(n^2-1) \sin(\phi) \sin(2\theta+\phi)}{\cos^2(\theta)+n^2\sin^2(\theta)}\right|.
    \label{eq:g0_norm}
\end{equation}
Here $n$ is the refractive index of SiN, $\phi$ is the phase shift due to the membrane thickness $d$, given by $\frac{nd\omega}{c}$ and $c$ the speed of light in vacuum. $\theta$ is the local phase of the resonant light, corresponding to the chip position. The relationship is expressed by $\theta=2\pi \frac{x}{\lambda}$. 

The optical losses (Fig.~\ref{fig: Optomechanical_spectra}b,d) caused by the slabs both display a periodicity of half-wavelength as well, consistent with those of the theoretical predictions for SM~\cite{jayich2008dispersive} and DM~\cite{Newsom2020}. Due to alignment imperfections of the chip normal to the incident light, the cavity resonance slightly shifts (c.f.\ Fig.~\ref{fig: Dispersion_curve}c) and the cavity linewidth lacks a distinct trend~\cite{lee2002spatial, sankey2010strong, karuza2012tunable}, which is why we refrain from fitting the data. We estimate that this misalignment contributes to the cavity loss by about \SI{763}{kHz}. Furthermore, we observe that the inter-membrane resonance shifts by up to \SI{0.1}{nm} when we move the membranes laterally with respect to the cavity axis. This effect indicates that the membranes are not perfectly parallel due to local variations in the substrate thickness. Nonetheless, the trends in $g_\mathrm{0}$ and $\kappa$ are similar, meaning that enhanced $g_\mathrm{0}$ also results in higher cavity loss. The cavity linewidth for both the SM and DMs cases tends toward the empty cavity linewidth, with a similar minimum measured value about \SI{213}{kHz}. The losses of DMs near the optimum coupling (near $0.25\lambda$ chip position) reach $1.17 \pm 0.03$ MHz, which is more than one order of magnitude higher than for the empty FP cavity. Note that the loss for a SM system also reaches $0.49 \pm 0.02$ MHz.

\begin{figure}
	\includegraphics[width = 0.44\textwidth]{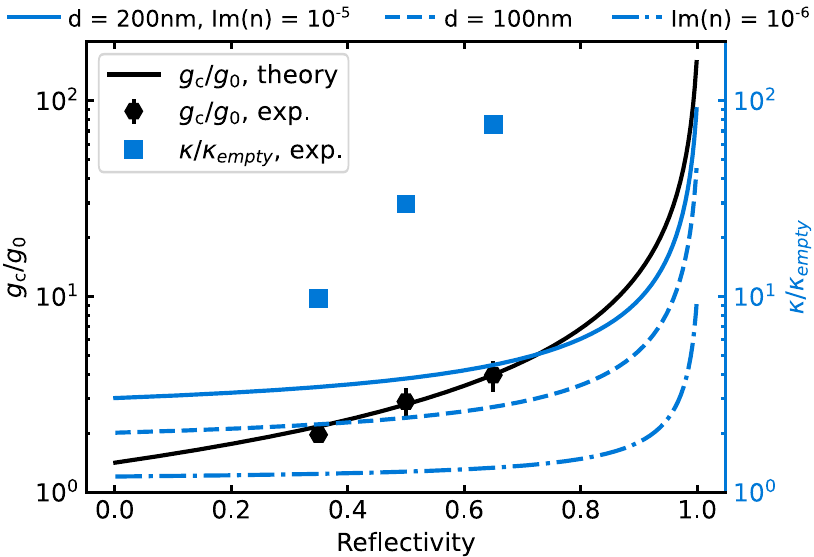}
	\caption{Enhancement of optomechanical coupling strength (black circles) and corresponding increase in cavity linewidth (blue squares) vs.\ membrane reflectivities. The black solid curve represents the enhancement in $g_\mathrm{c}$ and the blue curve illustrates the $\kappa_\mathrm{empty}$ and material absorption ($\kappa_\mathrm{abs}$) limited total cavity losses, applying adapted models from~\cite{Newsom2020} using our experimental parameters -- membrane thickness $d$ of \SI{200}{nm} and complex refractive index of $2+10^{-5}i$, where the imaginary part indicates absorption. The two dashed lines highlight potential improvements in $\kappa$ by thinning down $d$ to \SI{100}{nm} or reducing Im(n) to $10^{-6}$~\cite{karuza2012tunable}, respectively. Error bars of $g_\mathrm{c} /g_0$ are standard deviation derived from the fit.}
	\label{fig: enhancement}
\end{figure}

\section*{Discussion}
\label{sec: enhancement}

We have introduced a method to measure the enhancement of the single-photon optomechanical coupling rate, using long-range interactions in a multi-membrane system. We observe significant enhanced optomechanical coupling from a double-membrane device in a FP cavity when both the FP cavity and inter-membrane cavity resonance conditions are met. As shown in Fig.~\ref{fig: enhancement}, the enhancement of $g_\mathrm{c}$ of the collective motion matches theoretical predictions ~\cite{Xuereb2012, Newsom2020} and we observe enhancement of 1.97, 2.90, and 3.96 for membrane reflectivities of $R=35\%$, $R=50\%$, and $R=65\%$, respectively~\cite{Newsom2020}. The light couples to a hybridized long-range collective motion in which both the center-of-mass (COM) and breathing modes are present (see detailed analyses in Appendix), rather than only the latter alone dominating when enhancement occurs~\cite{Xuereb2012,xuereb2013collectively,Newsom2020}. Further enhancement would be possible with even higher reflectivity \cite{Li2016}, which is in principle readily available~\cite{Gaertner2018}. However, technical limitations in our ability to lock the laser to the cavity resonance currently prevents us from achieving higher coupling rates. Part of the challenge comes from imperfect alignment of the DMs inside the FP cavity, which results in high cavity losses \cite{sankey2010strong, karuza2012tunable}, which would be exacerbated even more when using higher reflectivities.

Our devices already feature ultra-low mechanical dissipation ($\gamma_\mathrm{M}\approx\SI{0.1}{Hz}$) but can be further improved by applying advanced mechanical engineering techniques~\cite{tsaturyan2017ultracoherent,enzian2023phononically, PhysRevX.14.011039}, which will directly allow to reach a regime where the optomechanical coupling rate is larger than the thermal decoherence rate in a cryogenic environment~\cite{planz2023membrane}. With the method demonstrated here of increasing $g_\mathrm{c}$, the main challenge to reach the single-photon strong coupling regime, where $g_\mathrm{c}>\kappa,\gamma_\mathrm{M}$, is to reduce the optical losses, caused by the FP cavity and the membranes inside. By improving the alignment between the FP cavity and membranes, it should be possible to significantly reduce scattering losses, leaving only material absorption and the empty cavity linewidth $\kappa_\mathrm{empty}$ (blue solid curve in Fig.~\ref{fig: enhancement}). The material absorption of membranes can be reduced by either thinning down the thickness or using even lower absorption material (dashed lines in Fig.~\ref{fig: enhancement}). For example, reducing the imaginary part of the refractive index of SiN to $10^{-6}$~\cite{karuza2012tunable} enhances $g_\mathrm{c}$ nearly tenfold relative to the increase in $\kappa$. Using silica instead of silicon nitride could further reduce the imaginary part by two orders of magnitude \cite{kitamura2007optical} and even lead to a narrowing of the optical linewidth~\cite{xuereb2013collectively}. At the same time, stable high-finesse FP cavities with only tens of kHz linewidth for cavities several millimeter long~\cite{PhysRevApplied.18.054054, Kelleher:23} have recently been realized. By shortening our cavity length to a few millimeters while preserving the long-range type of interaction ($L \gg \lambda_\mathrm{res}$), we can achieve an initial $g_\mathrm{0}$ on the order of hundreds of Hertz~\cite{rossi2018measurement,huang2024room}. For double-membranes with a reflectivity of $99.9\%$, we can therefore extrapolate that the enhancement could reach a factor of 157, which, with improved alignment and reduced losses, could allow us to get within the regime where $g_\mathrm{0}/\kappa \lesssim 1$, potentially reaching the single-photon strong coupling regime. Entering this regime will allow to observe novel effects, such as an optomechanical photon blockade~\cite{rabl2011photon} and the generation of non-Gaussian mechanical states~\cite{nunnenkamp2011single}.

Currently, the enhancement in $g_\mathrm{c}$ is comparable to the increase in $\kappa$, already leading to an enhancement of the single-photon cooperativity ($C_\mathrm{0}=4g_\mathrm{c}^2/\kappa\gamma_\mathrm{M}$)~\cite{Newsom2020}. Despite the higher losses, our results demonstrate a two-fold increase in $C_\mathrm{0}$, from $1.8\times10^{-4}$ to $3.9\times10^{-4}$, as $R$ goes from $35\%$ to $65\%$. Additionally, shortening the cavity length $L$ can directly increase $C_\mathrm{0}$ as both $g_\mathrm{0}$ and $\kappa$ scale as $1/L$~\cite{dumont2019flexure,enzian2023phononically,huang2024room}. Therefore, the increase in $g_\mathrm{c}$ can be used for enhanced optomechanical squeezing~\cite{safavi2013squeezed, nielsen2017multimode} and room-temperature quantum optomechanical experiments~\cite{huang2024room}.

\section*{Acknowledgements}
We would like thank David Vitali, Witlef Wieczorek, Menno Poot, and Mads Bjerregaard Kristensen for valuable discussions. We would like to thank the anonymous referees for reviewing our work and for their contributions to significantly improving the manuscript. We also acknowledge nanofabrication assistance from the Kavli Nanolab Delft and Jin Chang. This work is financially supported by the European Research Council (ERC CoG Q-ECHOS, 101001005) and is part of the research program of the Netherlands Organization for Scientific Research (NWO), supported by the NWO Frontiers of Nanoscience program, as well as through a Vrij Programma (680-92-18-04) grant.

\setcounter{figure}{0}
\renewcommand{\thefigure}{A\arabic{figure}}
\setcounter{equation}{0}
\renewcommand{\theequation}{A\arabic{equation}}
\setcounter{table}{0}
\renewcommand{\thetable}{A\arabic{table}}

\clearpage

\begin{center}
	\textsc{\Large Appendix} 
\end{center}
\label{Appendix}

\section*{Device characterization and setups}
\label{supp: deviceandsetup}

\begin{figure}[h]
	\centering
	\includegraphics[width = 0.46\textwidth]{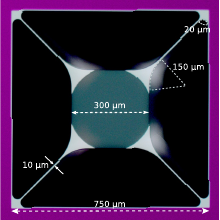}
	\caption{Highlights of design parameters of the SiN trampoline on a microscope image. The device is patterned over an area \SI{750}{\micro\meter} by \SI{750}{\micro\meter}. The width of the tether is \SI{10}{\micro\meter}. The membrane pattern is \SI{300}{\micro\meter} by \SI{300}{\micro\meter}. The inner fillet radius is \SI{150}{\micro\meter} and the outer one is \SI{20}{\micro\meter}, which reduces the stress concentration around corners~\cite{norte2016mechanical}. The photonic crystal pattern parameters are listed in Table~\ref{tab:dev}.}
	\label{fig: device_paras}
\end{figure}

The double-membrane (DM) devices are fabricated by following the same processes as in~\cite{Gaertner2018}. Parameters of the device patterns are illustrated in Fig.~\ref{fig: device_paras}. The front and back side patterns are aligned by using the same chip corner during the two electron beam lithography processes, which is necessary to pattern the devices on both sides of the same substrate. By carefully selecting the reference points, the misalignment can be minimized to below \SI{5}{\micro \meter}. The detailed parameters of three types of reflectivity devices are shown in Table~\ref{tab:dev}. The membrane's intensity reflection and transmission are first characterized in the setup described in~\cite{moura2018centimeter}. Then, the sample is loaded near the center in our high-finesse cavity setup~\cite{de2022coherent}. In detail, the cavity mirrors are mounted in a monolithic, stainless steel holder to keep their alignment and reduce their relative motion. One is mounted on top of a piezoelectric ring to control the cavity length. The sample holder is mounted on an $y$-$z$ alignment stage ($x$ being the cavity axis), which is mounted on a tip-tilt alignment stage. These all are placed in a vacuum chamber at pressures $<10^{-7}$~\si{\milli\bar} to minimize the viscous damping of the mechanics~\cite{bao2002energy}.

The double membranes form an inter-membrane cavity and the expected finesse $\mathcal{F}$ can be estimated by~\cite{stanley1994ultrahigh}
\begin{equation}
	\mathcal{F} = \frac{\pi\sqrt{R}}{1-R}.
\end{equation}
The inter-membrane cavity exhibits optical losses beyond the bare SiN material losses, with an extra round-trip loss exceeding $10^{-3}$, in addition to the external coupling due to transmission~\cite{Gaertner2018}. This indicates that the membranes introduce additional scattering and absorption losses when they are placed in the high-finesse cavity.

\begin{table}[!h]
	\centering
	\begin{tabular}{lccc}
		Device & \#1 & \#2 & \#3 \\
		\hline
		Reflectivity & 0.35 & 0.5 & 0.65 \\
		Lattice constant (nm)  & \SI{1240}{} & \SI{1310}{} & \SI{1340}{}\\
		Radius (nm) &  \SI{475}{}   & \SI{500}{} &\SI{514}{}\\
		Pad diameter (\SI{}{\micro m})&  \SI{300}{}   & \SI{300}{} &\SI{300}{}\\
		x-offset (\SI{}{\micro m})& \SI{41.11}{} & \SI{4.75}{} & \SI{33.52}{} \\
		y-offset (\SI{}{\micro m})& \SI{85.23}{} & \SI{2.75}{} & \SI{14.22}{} \\
		$\mathcal{F}$ @\SI{1550}{nm} & \SI{2.61}{} & \SI{4.00}{} & \SI{6.54}{} \\
		$\mathcal{F}$ (theory) & \SI{2.86}{} & \SI{4.44}{} & \SI{7.24}{} \\
		\hline
	\end{tabular}%
	\caption{Parameters of the 3 measured devices.}
	\label{tab:dev}%
\end{table}%

We drive our cavity with a laser beam originating from an ultra-low phase noise NKT Koheras Adjustik C15 with \SI{1}{\nano\meter} wavelength tunability centered around \SI{1550.12}{\nano\meter}. To stabilize the laser frequency to the cavity resonance, we utilize a Pound-Drever-Hall scheme~\cite{Black2001} with \SI{30}{\mega\hertz} sidebands. After reflecting from the cavity, part of the light is split off and subsequently detected on an avalanche photo diode, and this signal is mixed with another \SI{30}{\mega\hertz} tone derived from the same signal generator. The resulting error signal is fed to a proportional-integral-derivative (PID) controller that applies a modulation voltage to the laser. 

The rest of the reflected light from the cavity is sent to a 50-50 beam-splitter with a local oscillator driven by the same laser, and then detected using a home-built homodyne detector. A fiber-stretcher is used to stabilize the phase of the local oscillator.

\section*{Numerical model of Fabry-Pérot cavity with two membranes}
\label{FabryPerotmodel}

\begin{figure}[h!]
	\centering
	\includegraphics[width = 0.49\textwidth]{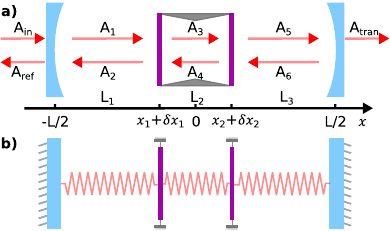}
	\caption{\textbf{a)} Schematic of TMM model for DMs inside a FP cavity. $\textbf{x}_\mathrm{1}$ and $\textbf{x}_\mathrm{2}$ are DMs static positions, respectively. When subjected to the radiation pressure, both membranes shift to new equilibrium positions, $\textbf{x}_\mathrm{1}+\delta \textbf{x}_\mathrm{1}$ and $\textbf{x}_\mathrm{2}+\delta \textbf{x}_\mathrm{2}$. In the dispersion relation simulations, this shift is not taken into account, as it only affects the inter-membrane cavity resonant wavelength in the dispersion relation. \textbf{b)} Optical spring model of coupled membranes through the radiation pressure inside the FP cavity. The strength of each spring depends on DMs positions ($\textbf{x}_\mathrm{1}$, $\delta \textbf{x}_\mathrm{1}$, $\textbf{x}_\mathrm{2}$, $\delta \textbf{x}_\mathrm{2}$). }
	\label{fig: TMM_mech}
\end{figure}

\begin{figure*}[t]
	\centering
	\includegraphics[width = 0.92\textwidth]{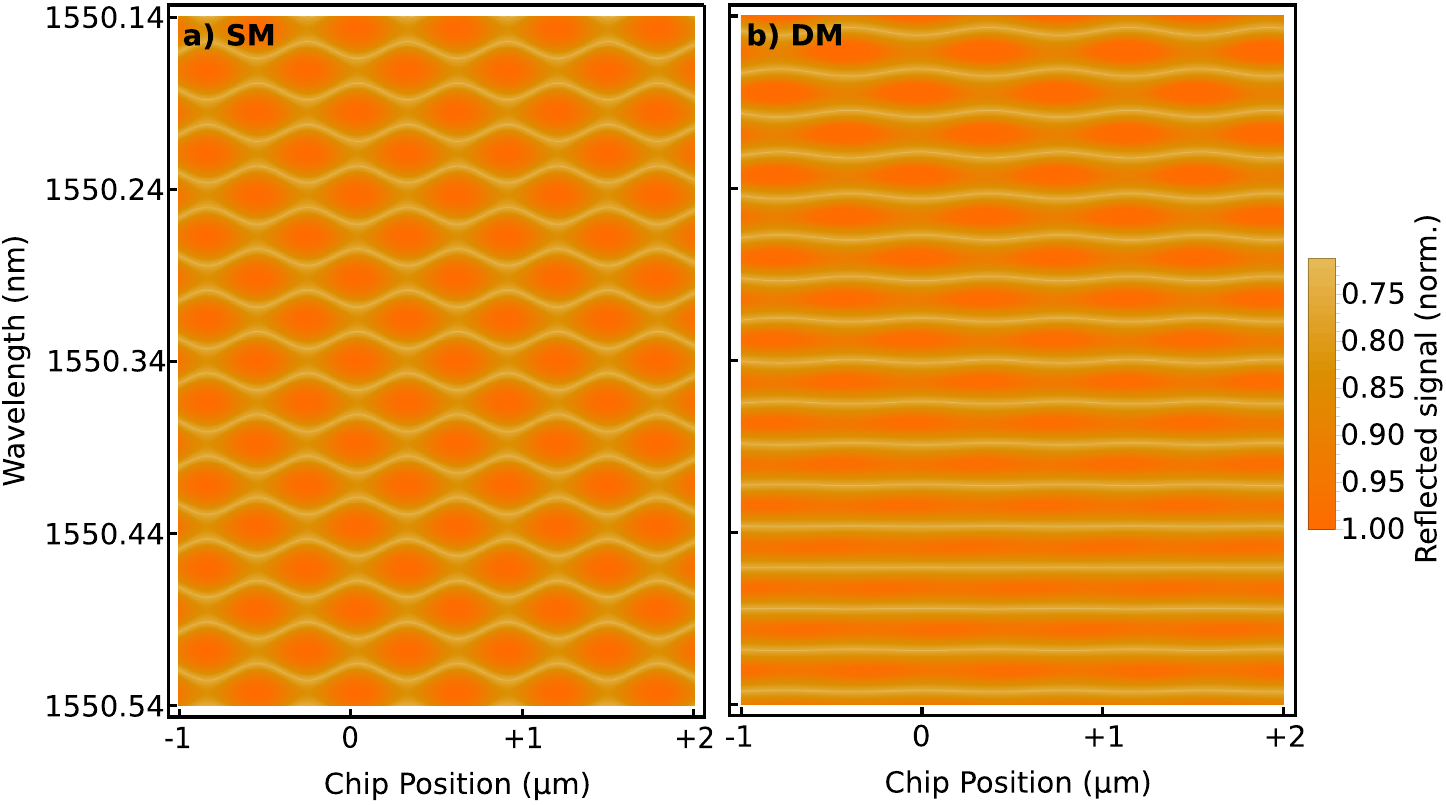}
	\caption{Simulated dispersion relation as a function of chip position and wavelength for a SM \textbf{a)} and DM \textbf{b)}. The reflected signal on resonance does not reach zero due to a trade-off between wavelength sweeping step size and computational costs. However, it still captures the resonance of the cavity.}
	\label{fig: disp_curves_SM_DM}
\end{figure*}

The optical properties of our system can be modelled by the transfer matrix method (TMM), by setting dielectric slabs inside a high-Finesse FP cavity~\cite{jayich2008dispersive, xuereb2013collectively, Li2016}. We simulate the dispersion relation by moving dielectric slabs along the cavity axis ($x$) (see Fig.~\ref{fig: TMM_mech}). The maximum slopes ($G = \mathrm{max}(|\partial \omega_\mathrm{c} /\partial x|)$) within different FP cavity free spectral ranges (FSR) of the single membrane (SM) are the same, and only depend on the reflectivity (see Fig.~\ref{fig: disp_curves_SM_DM}a). In contrast, $G$ of the double-membrane (DM) depends on both the wavelength and the membrane reflectivity (see \ Fig.~\ref{fig: disp_curves_SM_DM}b). Both cases give dispersion curves that are similar to the measured one in our experiments and in~\cite{Gaertner2018}.

Here, as illustrated in Fig.~\ref{fig: TMM_mech}a), we first describe the details of our DMs dispersion relations by TMM simulations, by applying expressions provided in \cite{Li2016}. The membrane's amplitude transmission and reflection coefficients are
\begin{equation}
	r_\mathrm{m} = \sqrt{R_\mathrm{m}},  \quad t_\mathrm{m} = \sqrt{T_\mathrm{m}},
\end{equation}
where $R_\mathrm{m}$, $T_\mathrm{m}$ are the intensity transmission and reflection coefficients of membranes, which are obtained from experiments. The reflection and transmission coefficients can be described by the material parameters of thin films~\cite{fan2002analysis, Li2016}
\begin{equation}
	\begin{aligned}
		& r_{\mathrm{m}}=\frac{\left(n^2-1\right) \sin \beta}{\left(n^2+1\right) \sin \beta+\mathrm{i} 2 n \cos \beta}, \\
		& t_{\mathrm{m}}=\frac{2 n}{\left(n^2+1\right) \sin \beta+\mathrm{i} 2 n \cos \beta},
	\end{aligned}
	\label{eq: rmtm}
\end{equation}
where $\beta=n k d$, and $k=2 \pi / \lambda$ is the wavenumber. This way $r_{\mathrm{m}}$ and $t_{\mathrm{m}}$ are complex, containing the phase shift of the light due to the membrane thickness $d$. The electric field amplitudes ($A_\mathrm{i},\ i = 1, ...\ , 6, \mathrm{ref}, \mathrm{tran}$) inside the cavity, transmitted, and reflected are given by:
\begin{equation}
	\begin{aligned}
		A_1 & =\mathrm{i} t A_{\mathrm{in}}+r A_2 \mathrm{e}^{\mathrm{i} k L_1}, \\
		A_2 & =\mathrm{i} t_{\mathrm{m}} A_4 \mathrm{e}^{\mathrm{i} k L_2}-r_{\mathrm{m}} A_1 \mathrm{e}^{\mathrm{i} k L_1}, \\
		A_3 & =\mathrm{i} t_{\mathrm{m}} A_1 \mathrm{e}^{\mathrm{i} k L_1}-r_{\mathrm{m}} A_4 \mathrm{e}^{\mathrm{i} k L_2}, \\
		A_4 & =\mathrm{i} t_{\mathrm{m}} A_6 \mathrm{e}^{\mathrm{i} k L_3}-r_{\mathrm{m}} A_3 \mathrm{e}^{\mathrm{i} k L_2}, \\
		A_5 & =\mathrm{i} t_{\mathrm{m}} A_3 \mathrm{e}^{\mathrm{i} k L_2}-r_{\mathrm{m}} A_6 \mathrm{e}^{\mathrm{i} k L_3}, \\
		A_6 & =r A_5 \mathrm{e}^{\mathrm{i} k L_3}, \\
		A_{\text {ref }} & =\mathrm{i} t A_2 \mathrm{e}^{\mathrm{i} k L_1}+r A_{\text {in }}, \\
		A_{\text {tran }} & =\mathrm{i} t A_5 \mathrm{e}^{\mathrm{i} k L_3},
	\end{aligned}
	\label{eq: TMM}
\end{equation}
where $r$ and $t$ are the amplitude reflection and transmission coefficients of our two identical FP cavity mirrors. 

\begin{figure}[h]
	\centering
	\includegraphics[width = 0.46\textwidth]{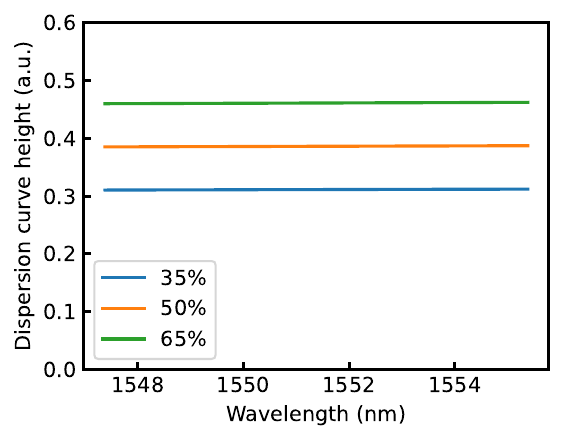}
	\caption{Normalized dispersion curve heights of three different reflectivity single membranes.}
	\label{fig: disp_single}
\end{figure}

\begin{figure*}[t]
	\centering
	\includegraphics[width = 0.9\textwidth]{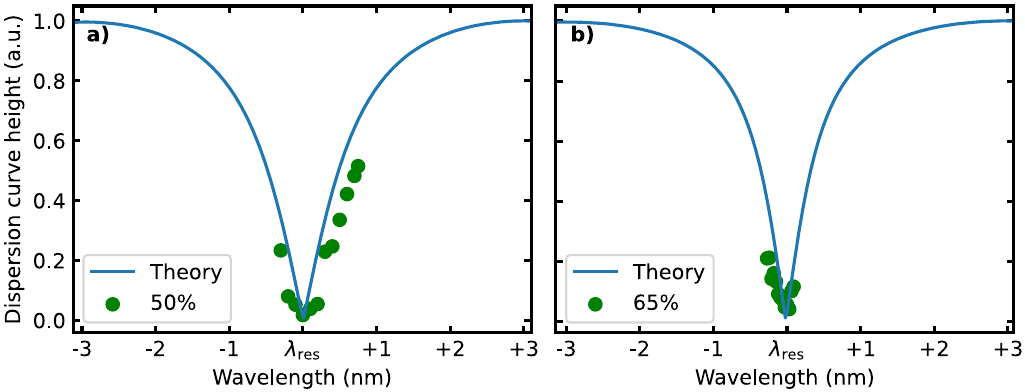}
	\caption{Dispersion curve height plots for the on-resonance devices of $R=50\%$ \textbf{a)} and $65\%$ \textbf{b)}. The measured resonance wavelength conditions are $\SI{1549.95}{nm}$ and \SI{1550.36}{nm}, respectively. }
	\label{fig: disp_heights}
\end{figure*}

In our simulation, we set $r = \sqrt{0.995}$ and $t=\sqrt{0.005}$, which results in a FP cavity linewidth of about \SI{32}{\femto\meter}, or \SI{4}{\mega\hertz}. This is in part due to the limited computational memory of our simulation tool. In practice, the $r_\mathrm{m}$ and $t_\mathrm{m}$ values differ from those of the bare-film for our devices of Eq.~\eqref{eq: rmtm} due to the photonic crystal patterned in the films~\cite{fan2002analysis}. The total cavity length is $L=\SI{50}{\milli\meter}$ and the membrane spacing $L_\mathrm{2} = \SI{200}{\micro\meter}$. By scanning the chip position and varying the wavelength, we obtain dispersion curves as shown in Fig.~\ref{fig: disp_curves_SM_DM}. The laser itself scans \SI{0.4}{\nano\meter} and covers $1/15$ of the FSR ($\sim\SI{6}{nm}$) of the inter-membrane cavity. Clearly, when approaching the resonance of the inter-membrane cavity, the dispersion curves become flat (c.f.\ Fig.~\ref{fig: disp_curves_SM_DM}b). Moreover, the on-resonant condition in our simulations indicates that the spacing between the membranes are integer multiple of half the cavity wavelength ($L_2 = n\lambda/2$) for both cases, with and without considering the phase shift due to dielectric membranes, which differs from the discussion in~\cite{Newsom2020}. By setting the reflectivity to \SI{50}{\%} and \SI{65}{\%} and extracting either the heights or the maximum slopes of the dispersion curves that are spaced by the high-finesse cavity FSR ($\sim\SI{24}{pm}$), we obtain the normalized dispersion curve heights (see Fig.~\ref{fig: disp_heights}). In contrast, it is constant for the single membrane case (see Fig.~\ref{fig: disp_single}). The linewidth of $\SI{65}{\%}$ is narrower than the one of $\SI{50}{\%}$ (blue curves), which gives a theoretical finesse of about $4.44$ and $6.54$ separately. The measured dispersion curve heights trace out a cavity resonance that is broadened (lower finesse) than predicted by our model, which can be attributed to the relative misalignment between the membranes.

\section*{Optomechanical coupling strength}

The optomechanical coupling strength $g_\mathrm{0}$ is evaluated from the mechanical spectra. We fit these spectra with the model provided in~\cite{de2022coherent}, which yields coupling rates for the individual membranes $g_\mathrm{0,j}$. We then compute the collective coupling $g_\mathrm{c}$ from the individual $g_\mathrm{0,j}$~\cite{Newsom2020}. For increased accuracy, we measure $g_\mathrm{0,j}$ at different powers and fit them using the same parameters. By repeating this procedure for different positions of the chip in the cavity, we experimentally obtain the position dependence $g_\mathrm{0}(x)$. 

The theoretical position dependence of $g_\mathrm{0}(x)$ is calculated by analyzing the light intensity across the membrane \cite{nielsen2017multimode, Newsom2020} for comparison. Here, we describe the fit model for the experimental spectra and the theory model for $g_\mathrm{0}$ separately.

\subsection*{Fit model for mechanical spectra}

Considering a regime where the optomechanical coupling strength is much smaller than the total cavity linewidth $\kappa$ (full width at half maximum), we use a linearized optomechanical formula, which describes a single cavity mode $\omega_\mathrm{c}$ interacting with two membranes' mechanical oscillation~\cite{de2022coherent}. A laser at frequency $\omega_\ell$ couples to the cavity with coupling strength $E = \sqrt{P_\ell \kappa_\mathrm{e}/\hbar\omega_\ell}$ with $P_\ell$ the laser power and $\kappa_\mathrm{e}$ the external coupling rate. This cavity contains two mechanical resonators at frequencies $\omega_{1,2} \simeq 2\pi\times 112$~\si{\kilo\hertz} with linewidths $\gamma_{1,2} \simeq 2\pi\times 0.1$~\si{\hertz}. These resonators are coupled with single-photon optomechanical coupling strengths $g_{0,1}$ and $g_{0,2}$ to the optical cavity. The Hamiltonian of this system \cite{de2022coherent} 
\begin{equation}
	\frac{\hat{H}}{\hbar}\!\!= \!\omega_\mathrm{c} \hat{a}^\dagger \hat{a} +\!\!\sum_{j = 1,2} \left(\frac{\omega_j}{2}\left( \hat{x}_j^2 \!+\! \hat{p}_j^2 \right)\!-\!g_{0,j}\hat{a}^\dagger \hat{a} \hat{x}_j \right) + iE\left( \hat{a}^\dagger e^{-i\omega_\ell t}\!\!-\!\mathrm{H.c.}\right)
	\label{Hamiltonian}
\end{equation}
with $\hat{a}$ ($\hat{a}^\dagger$) the annihilation (creation) operator of the optical mode, $\hat{x}_j$ and $\hat{p}_j$ the position and momentum operators of the two mechanical resonators ($j = 1,2$).The explicit formalism of $g_\mathrm{0}$ of a membrane inside a FP cavity is given in \cite{Cheung2011MIM}
\begin{equation}
	g_0 = x_\mathrm{zpf} \left(\left. \frac{\partial \omega_\mathrm{c}}{\partial x} \right|_{x=x_0} \right),
	\label{eq: g0_exact}
\end{equation}
where $x_\mathrm{zpf}=\sqrt{\hbar/2m_\mathrm{eff}\omega_\mathrm{M}}$ is the membrane eigenmode ($\omega_\mathrm{M}$) zero-point fluctuation, and $x_0$ is the rest position of the membrane. We coherently drive our cavity such that the cavity field has a large amplitude, $\lvert \langle \hat{a} \rangle\rvert \gg 1$, which allows us to separate the semi-classical averages and quantum fluctuations by rewriting the operators in Eq.~\eqref{Hamiltonian} as $\hat{O} = \langle \hat{O} \rangle + \delta \hat{O}$. By rotating the frame and including the coupling to the environment, we obtain the equations of motion for the fluctuations of mechanics and optical field

\begin{equation}
	\begin{aligned}
		\delta p_j &= -i \frac{\omega}{\omega_j} \delta x_j, \\
		\delta x_j &= \chi_j(\omega)\left[ G_j^* \delta a + G_j \delta a^\dagger + \xi_j \right], \\
		\delta a &= \chi_\mathrm{c}(\omega) \left( \sum_{j=1,2} i G_j \delta x_j + \sqrt{\kappa_\mathrm{e}} a^\mathrm{in} \right),
	\end{aligned}
	\label{eq: eom}
\end{equation}
with susceptibilities $\chi_j = \frac{\omega_j}{\omega_j^2 - \omega^2 - i\gamma_j \omega}$ and $\chi_\mathrm{c}(\omega) = \frac{1}{\kappa/2 + i(\Delta - \omega)}$.  Here $\gamma_j$ is the mechanical damping rate. $\hat{\xi}$ is the thermal noise driving term. For the mathematical details of the measured mechanical spectrum with homodyne detection scheme, we point to reference~\cite{de2022coherent}.

\subsection*{Coupling strength and the light intensity distribution}

\begin{figure}[h]
	\centering
	\includegraphics[width = 0.3\textwidth]{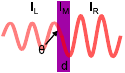}
	\caption{The intensity of the light field on the left ($I_\mathrm{L}$), inside the dielectric slab ($I_\mathrm{M}$), and on the right side ($I_\mathrm{R}$) depends on the local phase ($\theta$) of the resonant light. $d$ is the thickness of the slab, which gives rises to phase shift $\phi = \frac{nd\omega}{c}$ of light inside the slab.}
	\label{fig: light_intensity}
\end{figure}

The dielectric membrane is sensitive to the local phase ($\theta$) of the resonant light inside the cavity (see Fig.~\ref{fig: light_intensity}). Moving the device along the cavity axis ($z$-direction) will change the light field amplitudes at the either side of the membrane. Consequently, this will change the radiation pressure applied on the membranes~\cite{pedrotti2017introduction,Newsom2020}
\begin{equation}
	g_\mathrm{0} = x_\mathrm{zpf}\frac{\omega_\mathrm{c}}{L}N\left|I_\mathrm{R}-I_\mathrm{L}\right|,
	\label{eq: g0}
\end{equation}
where $N$ is a intensity normalization factor. The slab is thin compared to the cavity length ($d\ll L$) and $I_\mathrm{M}$ does not contribute to the radiation pressure, however does cause loss through its imaginary refractive index~\cite{Newsom2020}.

\begin{figure*}[t]
	\centering
	\includegraphics[width = 0.96\textwidth]{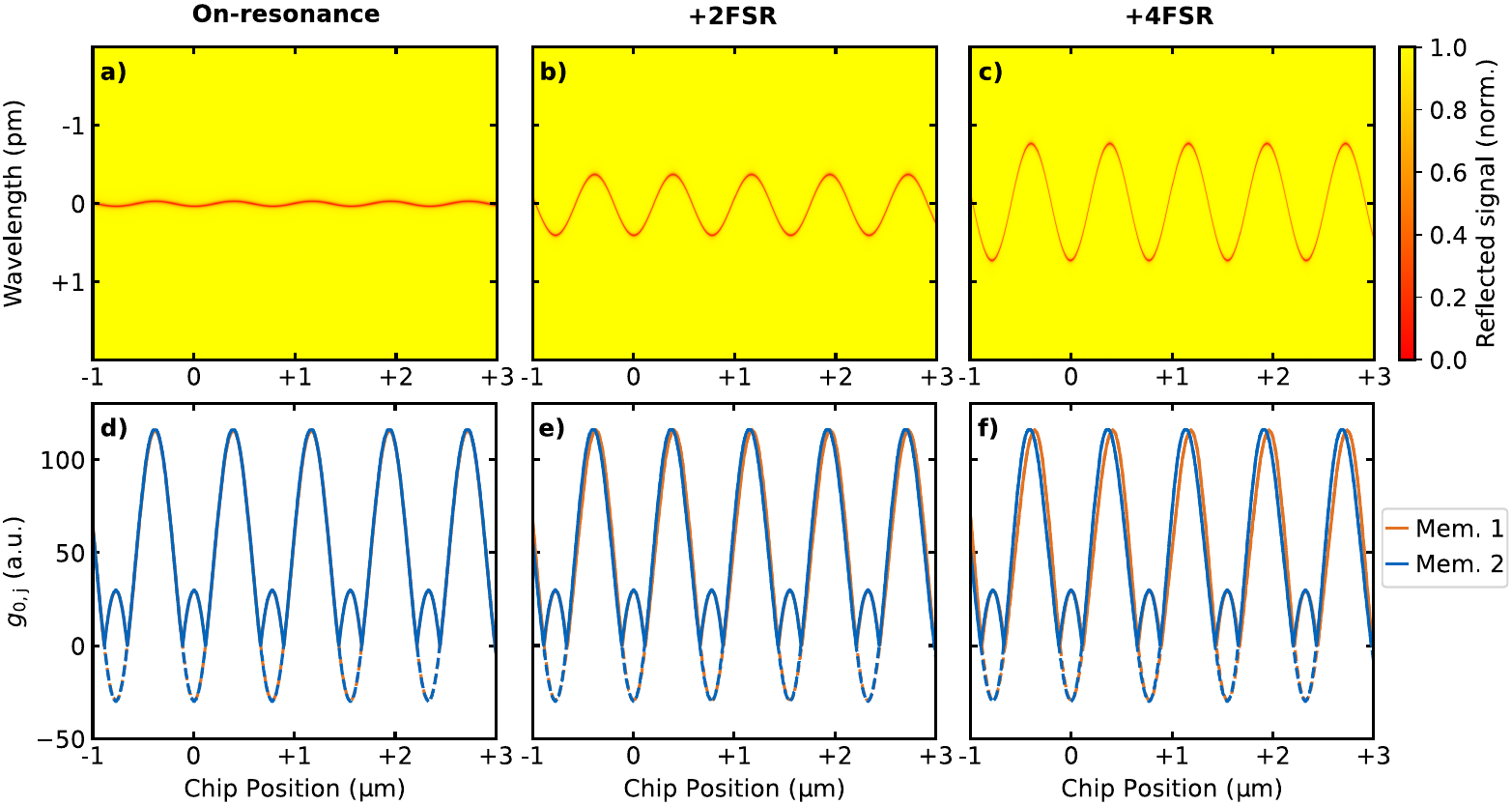}
	\caption{The top panels are zoom-ins of the dispersion curve simulations where the light is on-resonance with the inter-membrane cavity \textbf{a)}, two FP cavity FSR away \textbf{b)}, and four FP cavity FSR away \textbf{c)}, respectively. The wavelength step size is set to \SI{1}{fm}. The bottom panels \textbf{d)}, \textbf{e)}, and \textbf{f)} are the corresponding optomechanical coupling strength $g_\mathrm{0,j}$ of the two membranes, extracted at the minimum cavity reflected signal at all chip positions. The solid lines are evaluated with Eq.~\eqref{eq: intensity2}, while the dashed lines are without taking the absolute value in Eq.~\eqref{eq: intensity2}.}
	\label{fig: disp_g0}
\end{figure*}

The light field amplitudes on the left side, between, and on the right side of DMs can be obtained through TMM (Eq.~\eqref{eq: TMM}) simulation. First, we run simulations with a smaller step size ($\Delta\lambda \ll \kappa$) for three dispersion curves near the inter-membrane resonant wavelength (c.f.\ Fig.~\ref{fig: disp_g0}a-c). The dispersion curve height is much smaller than \SI{1}{pm} at the inter-membrane resonant wavelength. The corresponding light intensities are obtained by
\begin{equation}
	\begin{aligned}
		I_\mathrm{L} & =A_1^* A_1 + A_2^* A_2, \\
		I_\mathrm{inter} & =A_3^* A_3 + A_4^* A_4, \\
		I_\mathrm{R} & =A_5^* A_5 + A_6^* A_6.
	\end{aligned}
	\label{eq: intensity1}
\end{equation}
Then, the optomechanical coupling strength of two membranes can be evaluated through
\begin{equation}
	\begin{aligned}
		g_\mathrm{0,1} =  x_\mathrm{zpf}\frac{\omega_\mathrm{c}}{L} N_1 \left| I_\mathrm{inter}-I_\mathrm{L}\right|, \\
		g_\mathrm{0,2} = x_\mathrm{zpf}\frac{\omega_\mathrm{c}}{L} N_2 \left| I_\mathrm{R} - I_\mathrm{inter}\right|,
	\end{aligned}
	\label{eq: intensity2}
\end{equation}
where $N_\mathrm{1}$ and $N_\mathrm{2}$ are normalization factors that are given by 
\begin{equation}
	\begin{aligned}
		N_1 = \frac{1}{L_1 I_\mathrm{L} + L_2 I_\mathrm{inter}}, \\
		N_2 = \frac{1}{L_3 I_\mathrm{R} + L_2 I_\mathrm{inter}}.
	\end{aligned}
	\label{eq: intensity2_norm}
\end{equation}
The absolute value in Eqs.~\eqref{eq: intensity2} means $g_\mathrm{0,j}$ are non-negative. We can now obtain the collective optomechanical coupling strength $g_\mathrm{c}$ of the collective modes by \cite{Newsom2020}
\begin{equation}
	g_\mathrm{c} = \sqrt{g_\mathrm{0,1}^2 + g_\mathrm{0,2}^2}.
	\label{eq: intensity3}
\end{equation}
As shown in Fig.~\ref{fig: disp_g0}d at each chip position $g_\mathrm{0,1} = g_\mathrm{0,2}$ when the light is on-resonance with the inter-membrane cavity. The value of $g_\mathrm{0,j}$ and $g_\mathrm{c}$ is periodic with $\lambda_\mathrm{res}/2$. Furthermore, in each period, we can see that the light can be focusing outside the inter-membrane cavity (see dashed line in Fig.~\ref{fig: disp_g0}d), by the sign of $g_\mathrm{0,j}$ without taking the absolute value in Eq.~\eqref{eq: intensity2}. At these position, the $g_\mathrm{0,j}$ is much smaller than the maximum of $g_\mathrm{0,i}$, where the light is the highest between DM. Besides, moving the chip position slightly off the cavity center does not affect the maximum $g_\mathrm{0,j}$.

We perform the dispersion curve measurements at a wavelength step size of roughly \SI{100}{pm}, which is larger than the FP cavity FSR (\SI{24}{pm}). We further simulate the $g_\mathrm{0,j}$ when the light is detuned by \SI{2}{FSR} (see Fig.~\ref{fig: disp_g0}b, e) and by \SI{4}{FSR} (see Fig.~\ref{fig: disp_g0}c, f) away. Moving the laser wavelength off-resonance with the inter-membrane cavity has only a minor effect on the $g_\mathrm{0,j}$. Detuning it by 4 FSR off the main FP cavity (12 GHz or 96 pm) results in the shift shown in Fig.~\ref{fig: disp_g0}d-f.

To capture the light intensities distribution dependence on the chip position (or $\theta$) on either side (c.f.\ Fig.~\ref{fig: light_intensity}), we apply the transfer function~\cite{brooker2003modern} for the light across a dielectric slab
\begin{equation}
	\left(\begin{array}{c}
		\mathbf{E}_\mathrm{R}(\theta) \\
		Z_\mathrm{0} \mathbf{H}_\mathrm{R}(\theta)
	\end{array}\right)=\left(\begin{array}{cc}
		\cos (\phi) & -\frac{\mathrm{i}}{n} \sin (\phi) \\
		-\mathrm{i} n \sin (\phi) & \cos (\phi)
	\end{array}\right)\left(\begin{array}{c}
		\mathbf{E}_\mathrm{L}(\theta + \phi) \\
		Z_\mathrm{0} \mathbf{H}_\mathrm{L}(\theta + \phi)
	\end{array}\right),
	\label{eq: transfer_slab}
\end{equation} 
where $Z_\mathrm{0}=\sqrt{\mu_0 /\epsilon_0}$ is the impedance in the vacuum, $\phi = \frac{n \omega d}{c}$ is the phase shift due to the film thickness. $c$ is the speed of light in vacuum. $\mathbf{E}_\mathrm{R, L}$ and $\mathbf{H}_\mathrm{R, L}$ are electric field amplitude and magnetic field strength on the right (left) side of the membrane, respectively. Considering a plane wave that travels along the cavity axis ($z$) and only has one polarization, we can write electric field as 
\begin{equation}
	\mathbf{E}(z, t) = E_0 \sin(kz) \sin(\omega t),
	\label{eq: Efield}
\end{equation}
where $E_0$ is the electric field amplitude. Applying the relation $\nabla \times \mathbf{H}=\frac{\partial \epsilon_0 \mathbf{E}}{\partial t}$, we obtain that $\mathbf{H}$ satisfies
\begin{equation}
	\mathbf{H}(z, t) = \frac{\omega}{k} E_0 \cos(kz) \cos(\omega t).
	\label{eq: Hfield}
\end{equation}  
Inserting Eq.~\eqref{eq: Efield} and Eq.~\eqref{eq: Hfield} into Eq.~\eqref{eq: transfer_slab}, we obtain that the intensity $|\mathbf{E}|^2$ satisfies
\begin{equation}
	I_\mathrm{R} = I_\mathrm{L}\frac{\cos^2(\theta+\phi)+n^2\sin^2(\theta+\phi)}{\cos^2(\theta)+n^2\sin^2(\theta)} ,
	\label{eq: intensity}
\end{equation}
Inserting it into Eq.~\eqref{eq: g0} and normalizing relative to the maximum, we obtain the equation 
\begin{equation}
	g_\mathrm{0, norm} = \left|\frac{(n^2-1) \sin(\phi) \sin(2\theta+\phi)}{\cos^2(\theta)+n^2\sin^2(\theta)}\right|,
	\label{eq:g0_norm_App}
\end{equation}
which is the same as in~\cite{Newsom2020} and applies to different $R$ membranes. It shows that $g_\mathrm{0, norm}$ has a periodicity of half of the resonant wavelength.

\section*{Collective motion of double array}

Photons bounce back and forth inside the high-finesse FP cavity and mediate the interaction between the DMs. This leads to collective modes that are not localized on a single membrane~\cite{Xuereb2012}. Although our setup cannot directly measure the relative phase of the mechanical motion, unlike in~\cite{sheng2020self}, we can study the long-range correlations between DMs by analyzing the radiation pressure driven mechanics, supported by the dispersion curves. Prefacing that, the relative phases between two membranes split into a center-of-mass and breathing mode only due to the presence of a light field~\cite{xuereb2013collectively, Newsom2020, de2022coherent}. In contrast, thermal forces only drive the two membranes independently, without building any phase correlations. 

As shown in Fig.~\ref{fig: disp_curves_SM_DM}b and Fig.~\ref{fig: disp_g0}, the dispersion curves flatten when the light is near-resonant with both cavities. This leads to the light either being focused inside or outside the inter-membrane cavity, depending on the local phase of the light (c.f.\ Fig.~\ref{fig: disp_g0}d-f). Consequently, the effective radiation pressure on the two membranes have opposite sign ($F_\mathrm{opt} \propto -\nabla I(\mathbf{x})$), leading them to move in opposite directions. Moreover, if two membranes are identical, they oscillate out-of-phase, i.e.\ only couple to the breathing mode~\cite{Xuereb2012,xuereb2013collectively,Li2016,Newsom2020}.

\begin{figure}[th]
	\centering
	\includegraphics[width = 0.48\textwidth]{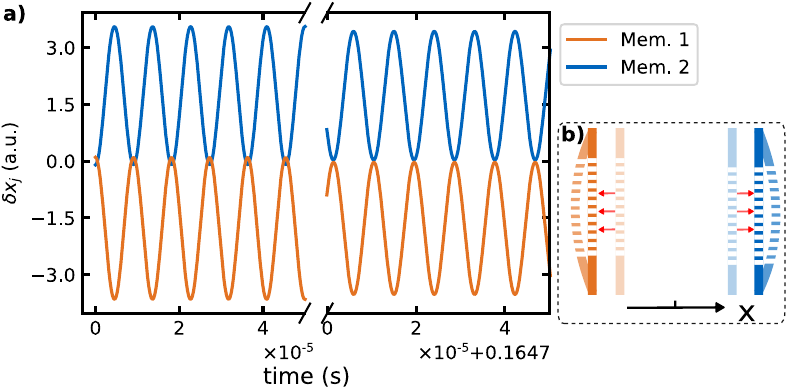}
	\caption{\text{a)} An exemplary plot of the time evolution of two identical membranes subjected to radiation pressure in the weak coupling regime, where $g_\mathbf{0}$ is maximally enhanced. We assume that the DMs start from slightly different initial positions, driven by thermal forces. \text{b)} Illustration of the optomechanical driven mechanical motion. The radiation pressure drives two membranes towards opposite equilibrium positions and results in out-of phase oscillation, i.e.\ breathing mode.}
	\label{fig: TMM_mech_timeevo}
\end{figure}

To establish a more precise framework, we model the two-membrane dynamics as coupled harmonic oscillators (c.f.\ Fig.~\ref{fig: TMM_mech}b)~\cite{POOT2012273, Matthijspra2023}, driven by radiation pressure that depends on the positions of the membranes and the input field $A_\mathrm{in}$. To do this, we replace $L_1$, $L_2$, and $L_3$ by $\mathbf{x}_1 + \delta \mathbf{x}_1 + L/2$, $\mathbf{x}_2 + \delta \mathbf{x}_2 - (\mathbf{x}_1 + \delta \mathbf{x}_1)$, and $L/2 - (\mathbf{x}_2 + \delta \mathbf{x}_2)$ in Eqs.~\eqref{eq: TMM}, respectively. Here, $\delta x_1$ and $\delta x_2$ are displacements induced by radiation pressure. Solving the cavity fields in each region allows us to obtain the radiation pressure force on each membrane, which is similar to Eqs.~\eqref{eq: intensity2} without the absolute value
\begin{equation}
	\begin{aligned}
		F_1(\mathbf{x}_1 + \delta \mathbf{x}_1, \mathbf{x}_2 + \delta \mathbf{x}_2) & \propto N_1 (I_\mathrm{L} - I_\mathrm{inter}), \\
		F_2(\mathbf{x}_1 + \delta \mathbf{x}_1, \mathbf{x}_2 + \delta \mathbf{x}_2) & \propto N_2 (I_\mathrm{inter} - I_\mathrm{R}).
	\end{aligned}
	\label{eq: radforce}
\end{equation}
In this manner, we derive the classical equations of motion of the DMs instead of relying on Eqs.~\eqref{eq: eom}. Inserting them into the two-membranes dynamics including the damping and driving terms we get
\begin{equation}
	\begin{aligned}
		\delta \ddot{\mathbf{x}}_1=-\omega_1^2 \delta \mathbf{x}_1-\gamma_1 \delta \dot{\mathbf{x}}_1+\frac{F_1\left(\mathbf{x}_1 + \delta \mathbf{x}_1, \mathbf{x}_2 + \delta \mathbf{x}_2\right)}{m_\mathrm{eff}}, \\
		\delta \ddot{\mathbf{x}}_2=-\omega_2^2 \delta \mathbf{x}_2-\gamma_2 \delta \dot{\mathbf{x}}_2+\frac{F_2\left(\mathbf{x}_1 + \delta \mathbf{x}_1, \mathbf{x}_2 + \delta \mathbf{x}_2\right)}{m_\mathrm{eff}},
	\end{aligned}
	\label{eq: TMM_eom}
\end{equation}
as $m_\mathrm{eff}$ is constant, we normalize it to one for simplicity. The exact expressions for $F_j(\mathbf{x}_1 + \delta \mathbf{x}_1, \mathbf{x}_2 + \delta \mathbf{x}_2)$ are too lengthy to be presented here, but they can be obtained by using Mathematica for example. Note that in our treatment, it is naturally assumed that the cavity field reacts instantaneously to the membrane's position, which requires the linewidths of both the main FP cavity and the inter-membrane cavity to be larger than the mechanical dynamics ($\kappa_\mathrm{e}, \kappa_\mathrm{DMs} \gg \omega_{j}$), which is the case in all of our experimental settings.

\begin{figure*}[t!]
	\centering
	\includegraphics[width = 0.97\textwidth]{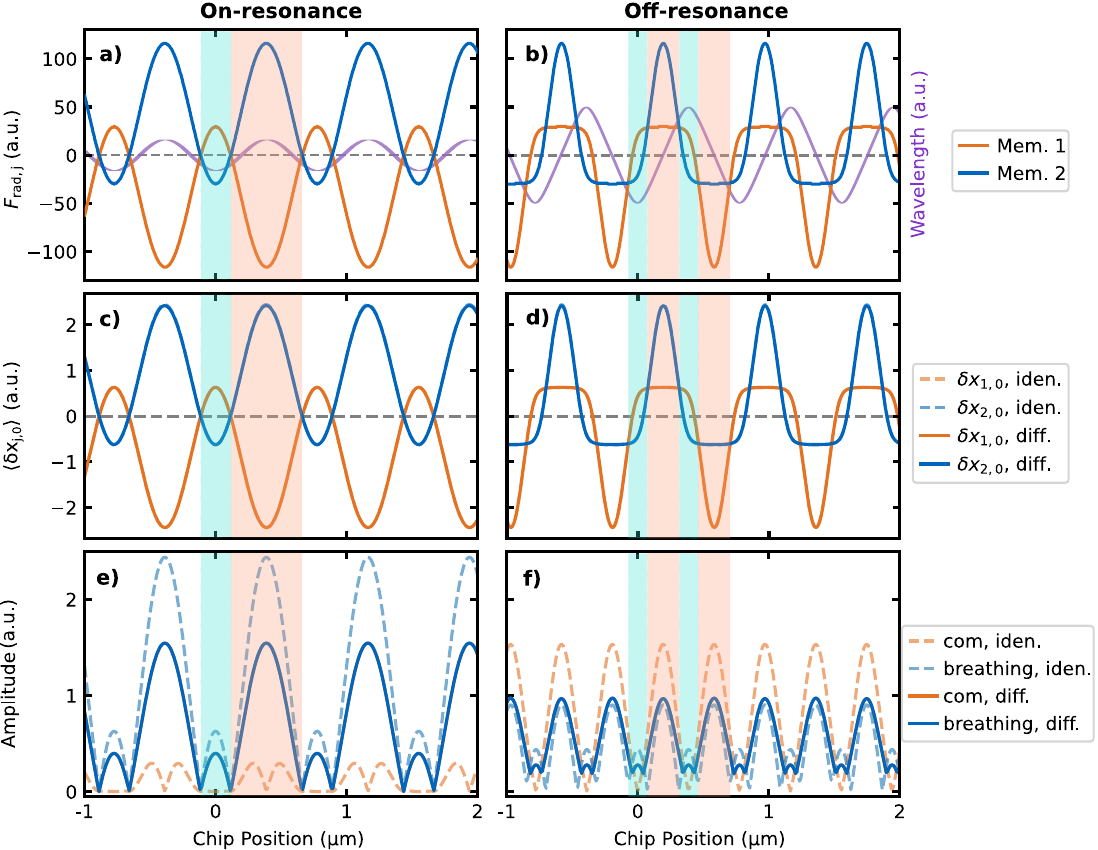}
	\caption{Two extreme cases of the effect of the radiation pressure on DMs mechanics in the weak coupling regime. Left and right panels correspond to on-resonant (flat dispersion curve) and off-resonant (the steepest dispersion curve) conditions, respectively. \textbf{a)} and \textbf{b)} Radiation pressure on each membranes. The purple curves represent the dispersion curves extracted only where the reflected signal is minimal at each chip positions (cf.\ Fig.~\ref{fig: disp_g0}). The wavelength axis is arbitrarily scaled to emphasize the correlation between radiation pressure and dispersion curves, without reflecting any steepness information. On-resonance, $F_\mathrm{rad,j}$ reaches the maximum when the dispersion is flat, whereas off-resonance, it is maximized when the dispersion curve slope is steepest. \textbf{c)} and \textbf{d)} The time-averaged displacement of each membranes due to radiation pressure. For both $\langle \delta x_{\mathrm{j,0}} \rangle$, dashed curves of two identical membranes are obscured by those of the non-identical membranes, respectively. The grey dashed horizontal lines in \textbf{a)}-\textbf{d)} highlight the original positions of each membrane without radiation pressure. \textbf{e)} and \textbf{f)} Averaged amplitudes of COM $ \left| \mathbf{Q} -\mathbf{Q}_0 \right| $  and breathing modes $\left| \mathbf{q} -\mathbf{q}_0 \right| $) in the orthogonal basis, extracted with $\overline{A}=\int_0^T A(t) d t$ and over a time interval $T$ that is significantly longer than both DMs mechanical oscillation period $2\pi/\omega_j$ and beating period $2\pi/ \left| \omega_1 - \omega_2 \right|$. In \textbf{c)}-\textbf{f)}, dashed and solid curves correspond to identical or different membranes, respectively, with the latter matching the devices in the main text. In the on-resonant panels, the light cyan and light orange shaded areas indicate regions where light is focused outside and between the membranes, respectively. Conversely, in the off-resonant panels, these shaded areas highlight the directions in which radiation pressure acts on the membranes, opposite for light cyan and the same for light orange. In \textbf{e)}, the solid orange curve is obscured by the solid blue one. In \textbf{f)}, part of dashed blue curve overlaps with the solid blue one, and the solid orange one is obscured by the blue one. The parameters used for these simulations are listed in Table~\ref{tab:parameters}.}
	\label{fig: TMM_com_bre}
\end{figure*}

The enhancement of the $g_0$ occurs when light is resonant with both the main FP and the inter-membrane cavities, i.e. on-resonance condition. This leads to the light field’s further localization between the membranes, and larger light intensity gradients across the membrane, resulting in larger radiation pressure and enhanced $g_0$. This phenomenon intrinsically occurs in the weak coupling regime, i.e. for small optical powers. Therefore, we set $A_\mathrm{in}$ to a level that compensates only for the mechanical damping rate $\gamma_j$ in Eqs.~\eqref{eq: TMM_eom}. By numerically solving these coupled optomechanical equations of motion, we obtain dynamics of the individual membranes. As shown in Fig.~\ref{fig: TMM_mech_timeevo}, when light is resonant with both cavities, two identical membranes are driven to opposite equilibrium positions and oscillate out-of-phase. This way, we can express the effect of radiation pressure as follows
\begin{equation}
	\delta \mathbf{x}_j = \langle \delta \mathbf{x}_{j,0} \rangle + \delta \mathbf{x}_j (\omega_j), \quad j = 1,2,
	\label{eq: rad_effect}
\end{equation}
where $\langle \delta \mathbf{x}_{j,0}\rangle = \frac{\omega_j}{2 \pi} \int_0^\frac{2\pi}{\omega_j}  \delta \mathbf{x}_j (t) dt$ is the time-averaged membrane's displacement, representing the radiation pressure induced displacement in the equilibrium position. $\delta \mathbf{x}_j (\omega_j)$ is the oscillation amplitude of individual membranes driven by radiation pressure at its eigenfrequencies~\cite{footnote2}. To study the collective behavior of the DMs, we introduce the COM ($\mathbf{Q}$) and breathing mode ($\mathbf{q}$) coordinates in the orthogonal basis~\cite{Li2016},
\begin{equation}
	\begin{aligned}
		\mathbf{Q} = \mathbf{Q}_0 + \frac{1}{2}[\delta \mathbf{x}_1 (\omega_1) + \delta \mathbf{x}_2 (\omega_2)],\\
		\mathbf{q} = \mathbf{q}_0 + \frac{1}{2}[ \delta \mathbf{x}_1 (\omega_1) - \delta \mathbf{x}_2 (\omega_2)],
	\end{aligned}
	\label{eq: TMM_collective_form}
\end{equation}
where $\mathbf{Q}_0$ and $\mathbf{q}_0$ are the time-averaged displacements of the COM and breathing mode, respectively, induced by radiation pressure. They are independent of time and frequency
\begin{equation}
	\begin{aligned}
		\mathbf{Q}_0 = \frac{1}{2}[\langle \delta \mathbf{x}_{1,0} \rangle +  \langle \delta \mathbf{x}_{2,0} \rangle],\\
		\mathbf{q}_0 = \frac{1}{2}[\langle \delta \mathbf{x}_{1,0} \rangle -  \langle \delta \mathbf{x}_{2,0} \rangle].
	\end{aligned}
	\label{eq: TMM_collective_form}
\end{equation}

As shown in Fig.~\ref{fig: TMM_com_bre}a, when light is resonant with both cavities, the radiation pressure on the DMs consistently exerts forces in opposite directions at various chip positions, in line with the behavior of $g_\mathrm{0}$ (c.f.\ Fig.~\ref{fig: disp_g0}). This leads to the DMs moving in the same direction as the radiation pressure (c.f.\ Fig.~\ref{fig: TMM_com_bre}c), resulting solely in $\mathbf{q}_0$ while $\mathbf{Q}_0$ remains zero, regardless of whether the DMs are identical or not. For DMs with identical eigenfrequencies, the breathing mode dominates across various chip positions (c.f.\ Fig.~\ref{fig: TMM_com_bre}e). Within the light orange shaded region, it is noteworthy that the COM also emerges and exhibits a derivative-dependent relationship with respect to radiation pressure. This arises from the strongly localized light field between the membranes, which generates a larger radiation pressure compared to the case where the light field is primarily distributed outside the inter-membrane cavity. Consequently, the mechanical oscillation amplitude becomes more sensitive to the gradient of the position-dependent radiation pressure. As a result, we expect the maximum COM amplitude to increase with the input optical power, $\left|A_\mathrm{in}\right|^2$, and to diminish as the light approaches the single-photon level. Nevertheless, light only couples to the breathing mode where $g_\mathrm{0}$ is maximally enhanced (c.f.\ Fig.~\ref{fig: TMM_com_bre}e), consistent with theoretical predictions~\cite{Xuereb2012,xuereb2013collectively,Newsom2020}. However, when the DMs possess different eigenfrequencies, due to fabrication imperfections for example, both COM and breathing modes are observed with equal amplitudes over multiple oscillation periods. 

Under the off-resonant condition where the dispersion curve is the steepest, radiation pressure on the DMs can act either in the same or opposite directions, depending on the local membrane positions (c.f.\ Fig.~\ref{fig: TMM_com_bre}b). When the radiation pressure acts in the same direction, light preferentially couples to one membrane, exerting a larger force on the left membrane in the reflective regime and a smaller force in the transmissive regime~\cite{xuereb2013collectively,stambaugh2015membrane}. This results in the DMs moving in the same direction (c.f.\ Fig.~\ref{fig: TMM_com_bre}d). For identical membranes, the breathing mode dominates in the cyan-shaded region, while both modes are present in the light orange region but the COM mode has a larger amplitude due to unequal radiation pressures. When the membranes have different eigenfrequencies, both the COM and breathing modes contribute equally over multiple oscillation periods, similar to the on-resonance condition.

\begin{figure}[h]
	\centering
	\includegraphics[width = 0.46\textwidth]{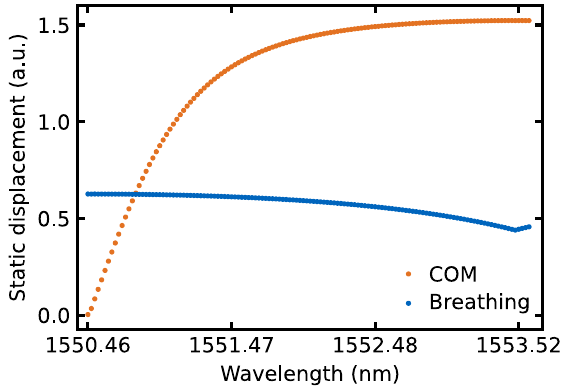}
	\caption{Maximum time-averaged COM and breathing modes displacement, across half of an inter-membrane FSR range ($\sim$3.06 nm). The dots represent the maximum displacement of COM ($\mathbf{Q}_0$) and breathing ($\mathbf{q}_0$) in every main FP cavity FSR range ($\sim$\SI{24.2}{pm}), respectively.}
	\label{fig: TMM_com_bre_static}
\end{figure}

As previously discussed, when the DMs have different eigenfrequencies as in our experiments, its collective motion exhibits both COM and breathing modes under both on-resonance and off-resonance conditions. However, the radiation pressure induced time-averaged displacement of the DMs is independent of the mechanical eigenfrequencies (c.f.\ Fig.~\ref{fig: TMM_com_bre}c and d). As shown in Fig.~\ref{fig: TMM_com_bre_static}, the $\mathbf{q}_0$ reaches its maximum under the on-resonance condition, with no $\mathbf{Q}_0$ present. When the light decouples from the inter-membrane cavity and approaches the region where the dispersion curve is the steepest, i.e.\ half the FSR of the inter-membrane cavity, the coupling to the $\mathbf{q}_0$ in each main FP cavity FSR decreases, while the coupling to the $\mathbf{Q}_0$ increases. 

\begin{table}[h]
	\begin{tabular}{cccc}
		\multicolumn{1}{l}{}       & Parameters       & \multicolumn{2}{c}{Values}    \\ \hline
		\multirow{7}{*}{Optics}    & $R$              & \multicolumn{2}{c}{0.995}     \\
		& $T$         & \multicolumn{2}{c}{0.005}     \\
		& $L$ (mm)                & \multicolumn{2}{c}{50}        \\
		& $L_2$ ($\mu$m) & \multicolumn{2}{c}{200}       \\
		& $d$ (nm)        & \multicolumn{2}{c}{200}       \\
		& $n$         & \multicolumn{2}{c}{$2 + 10^{-5} i$} \\
		& $A_\mathrm{in}$             & \multicolumn{2}{c}{0.01}      \\ \hline
		&      & Mem \#1        & Mem \#2      \\
		\multirow{3}{*}{Mechanics} & $\omega_j/2\pi$ (kHz) &   111.976   &  112.473            \\
		& $\gamma_j/2\pi$ (Hz)            &    0.1      & 0.1         \\
		& $m_\mathrm{eff}$                         & 1              & 1            \\ \hline
	\end{tabular}
	\caption{Summary of TMM numerical simulation parameters.}
	\label{tab:parameters}
\end{table}

\section*{Optomechanical coupling of double array}

\begin{figure}[h]
	\centering
	\includegraphics[width = 0.48\textwidth]{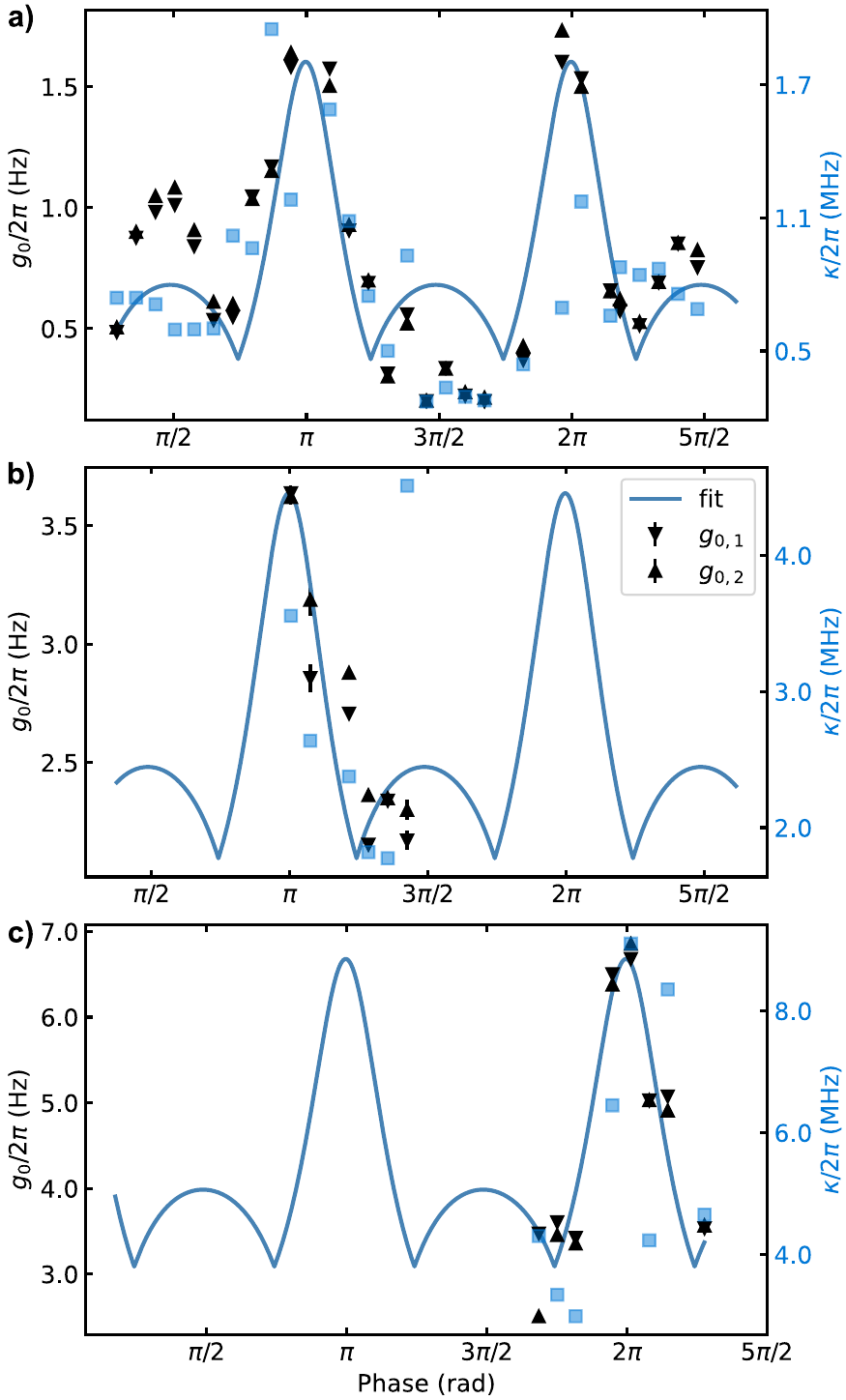}
	\caption{Optomechanical coupling strength $g_\mathrm{0,j}/2\pi$ (black triangles) and cavity linewidth $\kappa/2\pi$ (blue squares) of a) 35\%, b) 50\%, and c) 65\% reflective double membranes, respectively. The $g_\mathrm{0,j}/2\pi$ are fit by Eq.~\eqref{eq:g0_norm_App} (dark blue curves). The fits for $g_\mathrm{0,1}/2\pi$ and $g_\mathrm{0,2}/2\pi$ are very similar and therefore only $g_\mathrm{0,1}/2\pi$ is displayed. $\kappa/2\pi$ of all three cases are characterized at an input power of $\sim$\SI{10}{\micro W}.}
	\label{fig: g0_kappa_3R}
\end{figure}

In the main text, we present the collective coupling strengths $g_\mathrm{c}$ of the double membranes. Here, we provide both $g_\mathrm{0,1}$, $g_\mathrm{0,2}$ and $\kappa$ of all membranes that we investigated (see Fig.~\ref{fig: g0_kappa_3R} and Table~\ref{tab:g0}). All three devices with different $R$, $g_\mathrm{0,j}$ exhibit a clear dependence on $\theta$ (or the chip position) and all of them can be fit by Eq.~\eqref{eq:g0_norm_App}. However, the fit does not capture all the details of the $g_\mathrm{0,j}$. We attribute the imperfect fit in part to the noisy cavity locking, which could be improved by further increasing the stability of the setup. In addition, $g_\mathrm{0,j}$ near $3\pi/2$ is systematically lower than predicted. This discrepancy suggests the presence of additional coupling, which may be explained by the quadratic or quartic optomechanical coupling~\cite{sankey2010strong} or dissipative coupling~\cite{thompson2008strong,jayich2008dispersive,elste2009quantum,fitzgerald2021cavity,peralle2024quasibound,monsel2023dissipative}.

We can measure the reflected signal for low-$R$ at \SI{35}{\%} for the one-full wavelength period. However, for high-$R$, only $g_\mathrm{0,j}$ less than one-half period can be obtained. This limitation arises due to the shifting of the membrane splitting the cavity mode and the light is more confined in either sub-cavity, leading to a stronger reflected or transmitted signal~\cite{Xuereb2012, stambaugh2015membrane}. Our setup only measures the reflected signal, and therefore only one-half period of the signal can be measured for $R>0.5$. Additionally, a higher $R$ membrane introduces more scattering losses, resulting in a poor PDH error signal with broader and more shallow peaks. This limits the cavity locking and makes the characterization more difficult for higher reflective DMs. These challenges may be overcome by further optimizing the cavity and device alignment, and optimizing the configuration of devices, which does not have higher round-trip losses when the light circulates inside the high-finesse cavity. For example, a photonic crystal pattern designed for a Gaussian beam could decrease losses~\cite{guo2017integrated,Agrawal:24}.

\begin{table}[!h]
	\centering
	\begin{tabular}{cccc}
		Device & \#1 & \#2 & \#3 \\
		\hline
		Reflectivity & 0.35 & 0.5 & 0.65 \\
		$g_\mathrm{0}/2\pi$ (SM)  & $1.15\pm0.03$ & $1.77\pm0.26$ & $2.38\pm0.42$\\
		$g_\mathrm{0,1}/2\pi$ (DM) &  $1.60\pm0.05$   & $3.64\pm0.32$ &$6.68\pm0.01$\\
		$g_\mathrm{0,2}/2\pi$ (DM) &  $1.73\pm0.06$   & $3.62\pm0.32$ &$6.86\pm0.01$\\
		$g_\mathrm{c}/2\pi$ (DM) &  $2.27\pm0.07$   & $5.14\pm0.45$ &$9.45\pm0.02$\\
		$g_\mathrm{c}/g_\mathrm{0}$ &  $1.97$   & $2.90$ &$3.96$\\
		$\kappa/2\pi$ (DM) & $1173.6\pm33.3$ & $3555.6\pm27.5$ &  $9101.6\pm25.0$\\ 
		\hline
	\end{tabular}%
	\caption{$g_\mathrm{0}/2\pi$ (in Hz) and $\kappa/2\pi$ (in kHz) of SMs and DMs, respectively}
	\label{tab:g0}
\end{table}

\section*{Optomechanical coupling of a single membrane}

\begin{figure}[h]
	\centering
	\includegraphics[width = 0.48\textwidth]{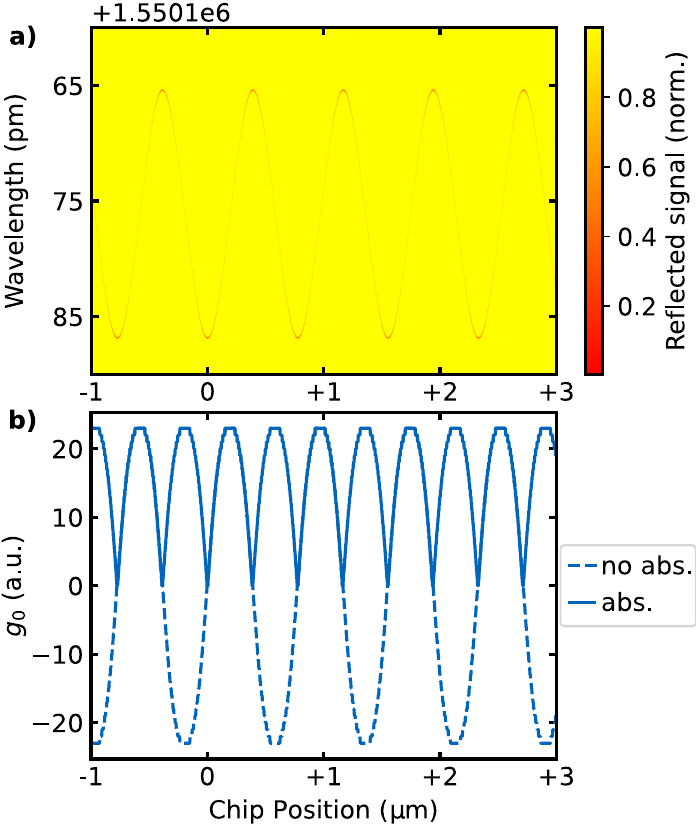}
	\caption{\textbf{a)} Zoom-in of dispersion curve simulation in one FSR of FP cavity of a SM. The wavelength step size is set to \SI{1}{fm}. \textbf{b)} The corresponding optomechanical coupling strength $g_\mathrm{0}$, extracted at the minimum cavity reflected signal at all chip positions. The solid lines are evaluated with Eq.~\eqref{eq: intensity2}, while the dashed lines are the ones without taking the absolute value in Eq.~\eqref{eq: intensity2}.}
	\label{fig: TMM_SM_dis_g0}
\end{figure}

In the main text, we also present the $g_\mathrm{0}$ of a single membrane. For completeness, we also provide the $g_\mathrm{0}$ obtained from the TMM simulation (see Fig.~\ref{fig: TMM_SM_dis_g0}). The dispersion curve exhibits a period of half-wavelength~\cite{thompson2008strong,Gaertner2018,Piergentili2018}, as expected. Correspondingly, the obtained $g_\mathrm{0}$ shows a quarter-wavelength periodicity, while having the same trend without taking the absolute value. We can use a $|\sin(\theta/2)|^2$ function to fit the $g_\mathrm{0}$, which we obtained from our experiments.

\begin{thebibliography}{70}%
	\makeatletter
	\providecommand \@ifxundefined [1]{%
		\@ifx{#1\undefined}
	}%
	\providecommand \@ifnum [1]{%
		\ifnum #1\expandafter \@firstoftwo
		\else \expandafter \@secondoftwo
		\fi
	}%
	\providecommand \@ifx [1]{%
		\ifx #1\expandafter \@firstoftwo
		\else \expandafter \@secondoftwo
		\fi
	}%
	\providecommand \natexlab [1]{#1}%
	\providecommand \enquote  [1]{``#1''}%
	\providecommand \bibnamefont  [1]{#1}%
	\providecommand \bibfnamefont [1]{#1}%
	\providecommand \citenamefont [1]{#1}%
	\providecommand \href@noop [0]{\@secondoftwo}%
	\providecommand \href [0]{\begingroup \@sanitize@url \@href}%
	\providecommand \@href[1]{\@@startlink{#1}\@@href}%
	\providecommand \@@href[1]{\endgroup#1\@@endlink}%
	\providecommand \@sanitize@url [0]{\catcode `\\12\catcode `\$12\catcode
		`\&12\catcode `\#12\catcode `\^12\catcode `\_12\catcode `\%12\relax}%
	\providecommand \@@startlink[1]{}%
	\providecommand \@@endlink[0]{}%
	\providecommand \url  [0]{\begingroup\@sanitize@url \@url }%
	\providecommand \@url [1]{\endgroup\@href {#1}{\urlprefix }}%
	\providecommand \urlprefix  [0]{URL }%
	\providecommand \Eprint [0]{\href }%
	\providecommand \doibase [0]{https://doi.org/}%
	\providecommand \selectlanguage [0]{\@gobble}%
	\providecommand \bibinfo  [0]{\@secondoftwo}%
	\providecommand \bibfield  [0]{\@secondoftwo}%
	\providecommand \translation [1]{[#1]}%
	\providecommand \BibitemOpen [0]{}%
	\providecommand \bibitemStop [0]{}%
	\providecommand \bibitemNoStop [0]{.\EOS\space}%
	\providecommand \EOS [0]{\spacefactor3000\relax}%
	\providecommand \BibitemShut  [1]{\csname bibitem#1\endcsname}%
	\let\auto@bib@innerbib\@empty
	\bibitem [{\citenamefont {Thompson}\ \emph {et~al.}(2008)\citenamefont
		{Thompson}, \citenamefont {Zwickl}, \citenamefont {Jayich}, \citenamefont
		{Marquardt}, \citenamefont {Girvin},\ and\ \citenamefont
		{Harris}}]{thompson2008strong}%
	\BibitemOpen
	\bibfield  {author} {\bibinfo {author} {\bibfnamefont {J.}~\bibnamefont
			{Thompson}}, \bibinfo {author} {\bibfnamefont {B.}~\bibnamefont {Zwickl}},
		\bibinfo {author} {\bibfnamefont {A.}~\bibnamefont {Jayich}}, \bibinfo
		{author} {\bibfnamefont {F.}~\bibnamefont {Marquardt}}, \bibinfo {author}
		{\bibfnamefont {S.}~\bibnamefont {Girvin}},\ and\ \bibinfo {author}
		{\bibfnamefont {J.}~\bibnamefont {Harris}},\ }\bibfield  {title} {\bibinfo
		{title} {Strong dispersive coupling of a high-finesse cavity to a
			micromechanical membrane},\ }\href {https://doi.org/10.1038/nature06715}
	{\bibfield  {journal} {\bibinfo  {journal} {Nature}\ }\textbf {\bibinfo
			{volume} {452}},\ \bibinfo {pages} {72} (\bibinfo {year} {2008})}\BibitemShut
	{NoStop}%
	\bibitem [{\citenamefont {Peterson}\ \emph {et~al.}(2016)\citenamefont
		{Peterson}, \citenamefont {Purdy}, \citenamefont {Kampel}, \citenamefont
		{Andrews}, \citenamefont {Yu}, \citenamefont {Lehnert},\ and\ \citenamefont
		{Regal}}]{Peterson2016PRL}%
	\BibitemOpen
	\bibfield  {author} {\bibinfo {author} {\bibfnamefont {R.~W.}\ \bibnamefont
			{Peterson}}, \bibinfo {author} {\bibfnamefont {T.~P.}\ \bibnamefont {Purdy}},
		\bibinfo {author} {\bibfnamefont {N.~S.}\ \bibnamefont {Kampel}}, \bibinfo
		{author} {\bibfnamefont {R.~W.}\ \bibnamefont {Andrews}}, \bibinfo {author}
		{\bibfnamefont {P.-L.}\ \bibnamefont {Yu}}, \bibinfo {author} {\bibfnamefont
			{K.~W.}\ \bibnamefont {Lehnert}},\ and\ \bibinfo {author} {\bibfnamefont
			{C.~A.}\ \bibnamefont {Regal}},\ }\bibfield  {title} {\bibinfo {title} {Laser
			cooling of a micromechanical membrane to the quantum backaction limit},\
	}\href {https://doi.org/10.1103/PhysRevLett.116.063601} {\bibfield  {journal}
		{\bibinfo  {journal} {Phys. Rev. Lett.}\ }\textbf {\bibinfo {volume} {116}},\
		\bibinfo {pages} {063601} (\bibinfo {year} {2016})}\BibitemShut {NoStop}%
	\bibitem [{\citenamefont {Reinhardt}\ \emph {et~al.}(2016)\citenamefont
		{Reinhardt}, \citenamefont {M{\"u}ller}, \citenamefont {Bourassa},\ and\
		\citenamefont {Sankey}}]{reinhardt2016ultralow}%
	\BibitemOpen
	\bibfield  {author} {\bibinfo {author} {\bibfnamefont {C.}~\bibnamefont
			{Reinhardt}}, \bibinfo {author} {\bibfnamefont {T.}~\bibnamefont
			{M{\"u}ller}}, \bibinfo {author} {\bibfnamefont {A.}~\bibnamefont
			{Bourassa}},\ and\ \bibinfo {author} {\bibfnamefont {J.~C.}\ \bibnamefont
			{Sankey}},\ }\bibfield  {title} {\bibinfo {title} {Ultralow-noise sin
			trampoline resonators for sensing and optomechanics},\ }\href
	{https://doi.org/https://doi.org/10.1103/PhysRevX.6.021001} {\bibfield
		{journal} {\bibinfo  {journal} {Phys. Rev. X.}\ }\textbf {\bibinfo {volume}
			{6}},\ \bibinfo {pages} {021001} (\bibinfo {year} {2016})}\BibitemShut
	{NoStop}%
	\bibitem [{\citenamefont {H{\"a}lg}\ \emph {et~al.}(2021)\citenamefont
		{H{\"a}lg}, \citenamefont {Gisler}, \citenamefont {Tsaturyan}, \citenamefont
		{Catalini}, \citenamefont {Grob}, \citenamefont {Krass}, \citenamefont
		{H{\'e}ritier}, \citenamefont {Mattiat}, \citenamefont {Thamm}, \citenamefont
		{Schirhagl} \emph {et~al.}}]{halg2021membrane}%
	\BibitemOpen
	\bibfield  {author} {\bibinfo {author} {\bibfnamefont {D.}~\bibnamefont
			{H{\"a}lg}}, \bibinfo {author} {\bibfnamefont {T.}~\bibnamefont {Gisler}},
		\bibinfo {author} {\bibfnamefont {Y.}~\bibnamefont {Tsaturyan}}, \bibinfo
		{author} {\bibfnamefont {L.}~\bibnamefont {Catalini}}, \bibinfo {author}
		{\bibfnamefont {U.}~\bibnamefont {Grob}}, \bibinfo {author} {\bibfnamefont
			{M.-D.}\ \bibnamefont {Krass}}, \bibinfo {author} {\bibfnamefont
			{M.}~\bibnamefont {H{\'e}ritier}}, \bibinfo {author} {\bibfnamefont
			{H.}~\bibnamefont {Mattiat}}, \bibinfo {author} {\bibfnamefont {A.-K.}\
			\bibnamefont {Thamm}}, \bibinfo {author} {\bibfnamefont {R.}~\bibnamefont
			{Schirhagl}}, \emph {et~al.},\ }\bibfield  {title} {\bibinfo {title}
		{Membrane-based scanning force microscopy},\ }\href
	{https://doi.org/https://doi.org/10.1103/PhysRevApplied.15.L021001}
	{\bibfield  {journal} {\bibinfo  {journal} {Phys. Rev. Appl.}\ }\textbf
		{\bibinfo {volume} {15}},\ \bibinfo {pages} {L021001} (\bibinfo {year}
		{2021})}\BibitemShut {NoStop}%
	\bibitem [{\citenamefont {Nielsen}\ \emph {et~al.}(2017)\citenamefont
		{Nielsen}, \citenamefont {Tsaturyan}, \citenamefont {M{\o}ller},
		\citenamefont {Polzik},\ and\ \citenamefont
		{Schliesser}}]{nielsen2017multimode}%
	\BibitemOpen
	\bibfield  {author} {\bibinfo {author} {\bibfnamefont {W.~H.~P.}\
			\bibnamefont {Nielsen}}, \bibinfo {author} {\bibfnamefont {Y.}~\bibnamefont
			{Tsaturyan}}, \bibinfo {author} {\bibfnamefont {C.~B.}\ \bibnamefont
			{M{\o}ller}}, \bibinfo {author} {\bibfnamefont {E.~S.}\ \bibnamefont
			{Polzik}},\ and\ \bibinfo {author} {\bibfnamefont {A.}~\bibnamefont
			{Schliesser}},\ }\bibfield  {title} {\bibinfo {title} {Multimode
			optomechanical system in the quantum regime},\ }\href
	{https://doi.org/10.1073/pnas.1608412114} {\bibfield  {journal} {\bibinfo
			{journal} {PNAS}\ }\textbf {\bibinfo {volume} {114}},\ \bibinfo {pages} {62}
		(\bibinfo {year} {2017})}\BibitemShut {NoStop}%
	\bibitem [{\citenamefont {Huang}\ \emph {et~al.}(2024)\citenamefont {Huang},
		\citenamefont {Beccari}, \citenamefont {Engelsen},\ and\ \citenamefont
		{Kippenberg}}]{huang2024room}%
	\BibitemOpen
	\bibfield  {author} {\bibinfo {author} {\bibfnamefont {G.}~\bibnamefont
			{Huang}}, \bibinfo {author} {\bibfnamefont {A.}~\bibnamefont {Beccari}},
		\bibinfo {author} {\bibfnamefont {N.~J.}\ \bibnamefont {Engelsen}},\ and\
		\bibinfo {author} {\bibfnamefont {T.~J.}\ \bibnamefont {Kippenberg}},\
	}\bibfield  {title} {\bibinfo {title} {Room-temperature quantum optomechanics
			using an ultralow noise cavity},\ }\href
	{https://doi.org/https://doi.org/10.1038/s41586-023-06997-3} {\bibfield
		{journal} {\bibinfo  {journal} {Nature}\ }\textbf {\bibinfo {volume} {626}},\
		\bibinfo {pages} {512} (\bibinfo {year} {2024})}\BibitemShut {NoStop}%
	\bibitem [{\citenamefont {Thomas}\ \emph {et~al.}(2021)\citenamefont {Thomas},
		\citenamefont {Parniak}, \citenamefont {{\O}stfeldt}, \citenamefont
		{M{\o}ller}, \citenamefont {B{\ae}rentsen}, \citenamefont {Tsaturyan},
		\citenamefont {Schliesser}, \citenamefont {Appel}, \citenamefont {Zeuthen},\
		and\ \citenamefont {Polzik}}]{thomas2021entanglement}%
	\BibitemOpen
	\bibfield  {author} {\bibinfo {author} {\bibfnamefont {R.~A.}\ \bibnamefont
			{Thomas}}, \bibinfo {author} {\bibfnamefont {M.}~\bibnamefont {Parniak}},
		\bibinfo {author} {\bibfnamefont {C.}~\bibnamefont {{\O}stfeldt}}, \bibinfo
		{author} {\bibfnamefont {C.~B.}\ \bibnamefont {M{\o}ller}}, \bibinfo {author}
		{\bibfnamefont {C.}~\bibnamefont {B{\ae}rentsen}}, \bibinfo {author}
		{\bibfnamefont {Y.}~\bibnamefont {Tsaturyan}}, \bibinfo {author}
		{\bibfnamefont {A.}~\bibnamefont {Schliesser}}, \bibinfo {author}
		{\bibfnamefont {J.}~\bibnamefont {Appel}}, \bibinfo {author} {\bibfnamefont
			{E.}~\bibnamefont {Zeuthen}},\ and\ \bibinfo {author} {\bibfnamefont {E.~S.}\
			\bibnamefont {Polzik}},\ }\bibfield  {title} {\bibinfo {title} {Entanglement
			between distant macroscopic mechanical and spin systems},\ }\href
	{https://doi.org/https://doi.org/10.1038/s41567-020-1031-5} {\bibfield
		{journal} {\bibinfo  {journal} {Nat. Phys.}\ }\textbf {\bibinfo {volume}
			{17}},\ \bibinfo {pages} {228} (\bibinfo {year} {2021})}\BibitemShut
	{NoStop}%
	\bibitem [{\citenamefont {Kristensen}\ \emph {et~al.}(2024)\citenamefont
		{Kristensen}, \citenamefont {Kralj}, \citenamefont {Langman},\ and\
		\citenamefont {Schliesser}}]{MadsPRL2024}%
	\BibitemOpen
	\bibfield  {author} {\bibinfo {author} {\bibfnamefont {M.~B.}\ \bibnamefont
			{Kristensen}}, \bibinfo {author} {\bibfnamefont {N.}~\bibnamefont {Kralj}},
		\bibinfo {author} {\bibfnamefont {E.~C.}\ \bibnamefont {Langman}},\ and\
		\bibinfo {author} {\bibfnamefont {A.}~\bibnamefont {Schliesser}},\ }\bibfield
	{title} {\bibinfo {title} {Long-lived and efficient optomechanical memory
			for light},\ }\href {https://doi.org/10.1103/PhysRevLett.132.100802}
	{\bibfield  {journal} {\bibinfo  {journal} {Phys. Rev. Lett.}\ }\textbf
		{\bibinfo {volume} {132}},\ \bibinfo {pages} {100802} (\bibinfo {year}
		{2024})}\BibitemShut {NoStop}%
	\bibitem [{\citenamefont {Xuereb}\ \emph {et~al.}(2012)\citenamefont {Xuereb},
		\citenamefont {Genes},\ and\ \citenamefont {Dantan}}]{Xuereb2012}%
	\BibitemOpen
	\bibfield  {author} {\bibinfo {author} {\bibfnamefont {A.}~\bibnamefont
			{Xuereb}}, \bibinfo {author} {\bibfnamefont {C.}~\bibnamefont {Genes}},\ and\
		\bibinfo {author} {\bibfnamefont {A.}~\bibnamefont {Dantan}},\ }\bibfield
	{title} {\bibinfo {title} {Strong coupling and long-range collective
			interactions in optomechanical arrays},\ }\href
	{https://doi.org/10.1103/PhysRevLett.109.223601} {\bibfield  {journal}
		{\bibinfo  {journal} {Phys. Rev. Lett.}\ }\textbf {\bibinfo {volume} {109}},\
		\bibinfo {pages} {223601} (\bibinfo {year} {2012})}\BibitemShut {NoStop}%
	\bibitem [{\citenamefont {Xuereb}\ \emph {et~al.}(2013)\citenamefont {Xuereb},
		\citenamefont {Genes},\ and\ \citenamefont
		{Dantan}}]{xuereb2013collectively}%
	\BibitemOpen
	\bibfield  {author} {\bibinfo {author} {\bibfnamefont {A.}~\bibnamefont
			{Xuereb}}, \bibinfo {author} {\bibfnamefont {C.}~\bibnamefont {Genes}},\ and\
		\bibinfo {author} {\bibfnamefont {A.}~\bibnamefont {Dantan}},\ }\bibfield
	{title} {\bibinfo {title} {Collectively enhanced optomechanical coupling in
			periodic arrays of scatterers},\ }\href
	{https://doi.org/10.1103/PhysRevA.88.053803} {\bibfield  {journal} {\bibinfo
			{journal} {Phys. Rev. A}\ }\textbf {\bibinfo {volume} {88}},\ \bibinfo
		{pages} {053803} (\bibinfo {year} {2013})}\BibitemShut {NoStop}%
	\bibitem [{\citenamefont {Xuereb}\ \emph {et~al.}(2014)\citenamefont {Xuereb},
		\citenamefont {Genes}, \citenamefont {Pupillo}, \citenamefont {Paternostro},\
		and\ \citenamefont {Dantan}}]{xuereb2014reconfigurable}%
	\BibitemOpen
	\bibfield  {author} {\bibinfo {author} {\bibfnamefont {A.}~\bibnamefont
			{Xuereb}}, \bibinfo {author} {\bibfnamefont {C.}~\bibnamefont {Genes}},
		\bibinfo {author} {\bibfnamefont {G.}~\bibnamefont {Pupillo}}, \bibinfo
		{author} {\bibfnamefont {M.}~\bibnamefont {Paternostro}},\ and\ \bibinfo
		{author} {\bibfnamefont {A.}~\bibnamefont {Dantan}},\ }\bibfield  {title}
	{\bibinfo {title} {Reconfigurable long-range phonon dynamics in
			optomechanical arrays},\ }\href
	{https://doi.org/10.1103/PhysRevLett.112.133604} {\bibfield  {journal}
		{\bibinfo  {journal} {Phys. Rev. Lett.}\ }\textbf {\bibinfo {volume} {112}},\
		\bibinfo {pages} {133604} (\bibinfo {year} {2014})}\BibitemShut {NoStop}%
	\bibitem [{\citenamefont {Shkarin}\ \emph {et~al.}(2014)\citenamefont
		{Shkarin}, \citenamefont {Flowers-Jacobs}, \citenamefont {Hoch},
		\citenamefont {Kashkanova}, \citenamefont {Deutsch}, \citenamefont
		{Reichel},\ and\ \citenamefont {Harris}}]{Shkarin2014}%
	\BibitemOpen
	\bibfield  {author} {\bibinfo {author} {\bibfnamefont {A.~B.}\ \bibnamefont
			{Shkarin}}, \bibinfo {author} {\bibfnamefont {N.~E.}\ \bibnamefont
			{Flowers-Jacobs}}, \bibinfo {author} {\bibfnamefont {S.~W.}\ \bibnamefont
			{Hoch}}, \bibinfo {author} {\bibfnamefont {A.~D.}\ \bibnamefont
			{Kashkanova}}, \bibinfo {author} {\bibfnamefont {C.}~\bibnamefont {Deutsch}},
		\bibinfo {author} {\bibfnamefont {J.}~\bibnamefont {Reichel}},\ and\ \bibinfo
		{author} {\bibfnamefont {J.~G.~E.}\ \bibnamefont {Harris}},\ }\bibfield
	{title} {\bibinfo {title} {Optically mediated hybridization between two
			mechanical modes},\ }\href {https://doi.org/10.1103/PhysRevLett.112.013602}
	{\bibfield  {journal} {\bibinfo  {journal} {Phys. Rev. Lett.}\ }\textbf
		{\bibinfo {volume} {112}},\ \bibinfo {pages} {013602} (\bibinfo {year}
		{2014})}\BibitemShut {NoStop}%
	\bibitem [{\citenamefont {Bemani}\ \emph {et~al.}(2017)\citenamefont {Bemani},
		\citenamefont {Motazedifard}, \citenamefont {Roknizadeh}, \citenamefont
		{Naderi},\ and\ \citenamefont {Vitali}}]{Bemani2017PRA}%
	\BibitemOpen
	\bibfield  {author} {\bibinfo {author} {\bibfnamefont {F.}~\bibnamefont
			{Bemani}}, \bibinfo {author} {\bibfnamefont {A.}~\bibnamefont
			{Motazedifard}}, \bibinfo {author} {\bibfnamefont {R.}~\bibnamefont
			{Roknizadeh}}, \bibinfo {author} {\bibfnamefont {M.~H.}\ \bibnamefont
			{Naderi}},\ and\ \bibinfo {author} {\bibfnamefont {D.}~\bibnamefont
			{Vitali}},\ }\bibfield  {title} {\bibinfo {title} {Synchronization dynamics
			of two nanomechanical membranes within a fabry-perot cavity},\ }\href
	{https://doi.org/10.1103/PhysRevA.96.023805} {\bibfield  {journal} {\bibinfo
			{journal} {Phys. Rev. A}\ }\textbf {\bibinfo {volume} {96}},\ \bibinfo
		{pages} {023805} (\bibinfo {year} {2017})}\BibitemShut {NoStop}%
	\bibitem [{\citenamefont {Sheng}\ \emph
		{et~al.}(2020{\natexlab{a}})\citenamefont {Sheng}, \citenamefont {Wei},
		\citenamefont {Yang},\ and\ \citenamefont {Wu}}]{Sheng2020}%
	\BibitemOpen
	\bibfield  {author} {\bibinfo {author} {\bibfnamefont {J.}~\bibnamefont
			{Sheng}}, \bibinfo {author} {\bibfnamefont {X.}~\bibnamefont {Wei}}, \bibinfo
		{author} {\bibfnamefont {C.}~\bibnamefont {Yang}},\ and\ \bibinfo {author}
		{\bibfnamefont {H.}~\bibnamefont {Wu}},\ }\bibfield  {title} {\bibinfo
		{title} {Self-organized synchronization of phonon lasers},\ }\href
	{https://doi.org/10.1103/PhysRevLett.124.053604} {\bibfield  {journal}
		{\bibinfo  {journal} {Phys. Rev. Lett.}\ }\textbf {\bibinfo {volume} {124}},\
		\bibinfo {pages} {053604} (\bibinfo {year} {2020}{\natexlab{a}})}\BibitemShut
	{NoStop}%
	\bibitem [{\citenamefont {Xu}\ \emph {et~al.}(2016)\citenamefont {Xu},
		\citenamefont {Mason}, \citenamefont {Jiang},\ and\ \citenamefont
		{Harris}}]{Xu2016}%
	\BibitemOpen
	\bibfield  {author} {\bibinfo {author} {\bibfnamefont {H.}~\bibnamefont
			{Xu}}, \bibinfo {author} {\bibfnamefont {D.}~\bibnamefont {Mason}}, \bibinfo
		{author} {\bibfnamefont {L.}~\bibnamefont {Jiang}},\ and\ \bibinfo {author}
		{\bibfnamefont {J.~G.~E.}\ \bibnamefont {Harris}},\ }\bibfield  {title}
	{\bibinfo {title} {Topological energy transfer in an optomechanical system
			with exceptional points},\ }\href {https://doi.org/10.1038/nature18604}
	{\bibfield  {journal} {\bibinfo  {journal} {Nature}\ }\textbf {\bibinfo
			{volume} {537}},\ \bibinfo {pages} {80} (\bibinfo {year} {2016})}\BibitemShut
	{NoStop}%
	\bibitem [{\citenamefont {Yang}\ \emph {et~al.}(2020)\citenamefont {Yang},
		\citenamefont {Wei}, \citenamefont {Sheng},\ and\ \citenamefont
		{Wu}}]{yang2020phonon}%
	\BibitemOpen
	\bibfield  {author} {\bibinfo {author} {\bibfnamefont {C.}~\bibnamefont
			{Yang}}, \bibinfo {author} {\bibfnamefont {X.}~\bibnamefont {Wei}}, \bibinfo
		{author} {\bibfnamefont {J.}~\bibnamefont {Sheng}},\ and\ \bibinfo {author}
		{\bibfnamefont {H.}~\bibnamefont {Wu}},\ }\bibfield  {title} {\bibinfo
		{title} {Phonon heat transport in cavity-mediated optomechanical
			nanoresonators},\ }\href
	{https://doi.org/https://doi.org/10.1038/s41467-020-18426-4} {\bibfield
		{journal} {\bibinfo  {journal} {Nat. Commun.}\ }\textbf {\bibinfo {volume}
			{11}},\ \bibinfo {pages} {4656} (\bibinfo {year} {2020})}\BibitemShut
	{NoStop}%
	\bibitem [{\citenamefont {Weaver}\ \emph {et~al.}(2017)\citenamefont {Weaver},
		\citenamefont {Buters}, \citenamefont {Luna}, \citenamefont {Eerkens},
		\citenamefont {Heeck}, \citenamefont {de~Man},\ and\ \citenamefont
		{Bouwmeester}}]{Weaver2017}%
	\BibitemOpen
	\bibfield  {author} {\bibinfo {author} {\bibfnamefont {M.~J.}\ \bibnamefont
			{Weaver}}, \bibinfo {author} {\bibfnamefont {F.}~\bibnamefont {Buters}},
		\bibinfo {author} {\bibfnamefont {F.}~\bibnamefont {Luna}}, \bibinfo {author}
		{\bibfnamefont {H.}~\bibnamefont {Eerkens}}, \bibinfo {author} {\bibfnamefont
			{K.}~\bibnamefont {Heeck}}, \bibinfo {author} {\bibfnamefont
			{S.}~\bibnamefont {de~Man}},\ and\ \bibinfo {author} {\bibfnamefont
			{D.}~\bibnamefont {Bouwmeester}},\ }\bibfield  {title} {\bibinfo {title}
		{Coherent optomechanical state transfer between disparate mechanical
			resonators},\ }\href
	{https://doi.org/https://doi.org/10.1038/s41467-017-00968-9} {\bibfield
		{journal} {\bibinfo  {journal} {Nat. Commun.}\ }\textbf {\bibinfo {volume}
			{8}},\ \bibinfo {pages} {824} (\bibinfo {year} {2017})}\BibitemShut {NoStop}%
	\bibitem [{\citenamefont {de~Jong}\ \emph {et~al.}(2022)\citenamefont
		{de~Jong}, \citenamefont {Li}, \citenamefont {G{\"a}rtner}, \citenamefont
		{Norte},\ and\ \citenamefont {Gr{\"o}blacher}}]{de2022coherent}%
	\BibitemOpen
	\bibfield  {author} {\bibinfo {author} {\bibfnamefont {M.~H.}\ \bibnamefont
			{de~Jong}}, \bibinfo {author} {\bibfnamefont {J.}~\bibnamefont {Li}},
		\bibinfo {author} {\bibfnamefont {C.}~\bibnamefont {G{\"a}rtner}}, \bibinfo
		{author} {\bibfnamefont {R.~A.}\ \bibnamefont {Norte}},\ and\ \bibinfo
		{author} {\bibfnamefont {S.}~\bibnamefont {Gr{\"o}blacher}},\ }\bibfield
	{title} {\bibinfo {title} {Coherent mechanical noise cancellation and
			cooperativity competition in optomechanical arrays},\ }\href
	{https://doi.org/10.1364/OPTICA.446434} {\bibfield  {journal} {\bibinfo
			{journal} {Optica}\ }\textbf {\bibinfo {volume} {9}},\ \bibinfo {pages} {170}
		(\bibinfo {year} {2022})}\BibitemShut {NoStop}%
	\bibitem [{\citenamefont {Marzioni}\ \emph {et~al.}(2023)\citenamefont
		{Marzioni}, \citenamefont {Rasponi}, \citenamefont {Piergentili},
		\citenamefont {Natali}, \citenamefont {Di~Giuseppe},\ and\ \citenamefont
		{Vitali}}]{Marzioni2023}%
	\BibitemOpen
	\bibfield  {author} {\bibinfo {author} {\bibfnamefont {F.}~\bibnamefont
			{Marzioni}}, \bibinfo {author} {\bibfnamefont {F.}~\bibnamefont {Rasponi}},
		\bibinfo {author} {\bibfnamefont {P.}~\bibnamefont {Piergentili}}, \bibinfo
		{author} {\bibfnamefont {R.}~\bibnamefont {Natali}}, \bibinfo {author}
		{\bibfnamefont {G.}~\bibnamefont {Di~Giuseppe}},\ and\ \bibinfo {author}
		{\bibfnamefont {D.}~\bibnamefont {Vitali}},\ }\bibfield  {title} {\bibinfo
		{title} {Amplitude and phase noise in two-membrane cavity optomechanics},\
	}\bibfield  {journal} {\bibinfo  {journal} {Front. Phys.}\ }\textbf {\bibinfo
		{volume} {11}},\ \href {https://doi.org/10.3389/fphy.2023.1222056}
	{10.3389/fphy.2023.1222056} (\bibinfo {year} {2023})\BibitemShut {NoStop}%
	\bibitem [{\citenamefont {Aspelmeyer}\ \emph {et~al.}(2014)\citenamefont
		{Aspelmeyer}, \citenamefont {Kippenberg},\ and\ \citenamefont
		{Marquardt}}]{Aspelmeyer2014}%
	\BibitemOpen
	\bibfield  {author} {\bibinfo {author} {\bibfnamefont {M.}~\bibnamefont
			{Aspelmeyer}}, \bibinfo {author} {\bibfnamefont {T.~J.}\ \bibnamefont
			{Kippenberg}},\ and\ \bibinfo {author} {\bibfnamefont {F.}~\bibnamefont
			{Marquardt}},\ }\bibfield  {title} {\bibinfo {title} {Cavity optomechanics},\
	}\href {https://doi.org/10.1103/RevModPhys.86.1391} {\bibfield  {journal}
		{\bibinfo  {journal} {Rev. Mod. Phys.}\ }\textbf {\bibinfo {volume} {86}},\
		\bibinfo {pages} {1391} (\bibinfo {year} {2014})}\BibitemShut {NoStop}%
	\bibitem [{\citenamefont {Bowen}\ and\ \citenamefont
		{Milburn}(2015)}]{bowen2015quantum}%
	\BibitemOpen
	\bibfield  {author} {\bibinfo {author} {\bibfnamefont {W.~P.}\ \bibnamefont
			{Bowen}}\ and\ \bibinfo {author} {\bibfnamefont {G.~J.}\ \bibnamefont
			{Milburn}},\ }\href@noop {} {\emph {\bibinfo {title} {Quantum
				optomechanics}}}\ (\bibinfo  {publisher} {CRC press},\ \bibinfo {year}
	{2015})\BibitemShut {NoStop}%
	\bibitem [{\citenamefont {Rabl}(2011)}]{rabl2011photon}%
	\BibitemOpen
	\bibfield  {author} {\bibinfo {author} {\bibfnamefont {P.}~\bibnamefont
			{Rabl}},\ }\bibfield  {title} {\bibinfo {title} {Photon blockade effect in
			optomechanical systems},\ }\href
	{https://doi.org/https://doi.org/10.1103/PhysRevLett.107.063601} {\bibfield
		{journal} {\bibinfo  {journal} {Phys. Rev. Lett.}\ }\textbf {\bibinfo
			{volume} {107}},\ \bibinfo {pages} {063601} (\bibinfo {year}
		{2011})}\BibitemShut {NoStop}%
	\bibitem [{\citenamefont {Nunnenkamp}\ \emph {et~al.}(2011)\citenamefont
		{Nunnenkamp}, \citenamefont {B{\o}rkje},\ and\ \citenamefont
		{Girvin}}]{nunnenkamp2011single}%
	\BibitemOpen
	\bibfield  {author} {\bibinfo {author} {\bibfnamefont {A.}~\bibnamefont
			{Nunnenkamp}}, \bibinfo {author} {\bibfnamefont {K.}~\bibnamefont
			{B{\o}rkje}},\ and\ \bibinfo {author} {\bibfnamefont {S.~M.}\ \bibnamefont
			{Girvin}},\ }\bibfield  {title} {\bibinfo {title} {Single-photon
			optomechanics},\ }\href
	{https://doi.org/https://doi.org/10.1103/PhysRevLett.107.063602} {\bibfield
		{journal} {\bibinfo  {journal} {Phys. Rev. Lett.}\ }\textbf {\bibinfo
			{volume} {107}},\ \bibinfo {pages} {063602} (\bibinfo {year}
		{2011})}\BibitemShut {NoStop}%
	\bibitem [{\citenamefont {Safavi-Naeini}\ \emph {et~al.}(2013)\citenamefont
		{Safavi-Naeini}, \citenamefont {Gr{\"o}blacher}, \citenamefont {Hill},
		\citenamefont {Chan}, \citenamefont {Aspelmeyer},\ and\ \citenamefont
		{Painter}}]{safavi2013squeezed}%
	\BibitemOpen
	\bibfield  {author} {\bibinfo {author} {\bibfnamefont {A.~H.}\ \bibnamefont
			{Safavi-Naeini}}, \bibinfo {author} {\bibfnamefont {S.}~\bibnamefont
			{Gr{\"o}blacher}}, \bibinfo {author} {\bibfnamefont {J.~T.}\ \bibnamefont
			{Hill}}, \bibinfo {author} {\bibfnamefont {J.}~\bibnamefont {Chan}}, \bibinfo
		{author} {\bibfnamefont {M.}~\bibnamefont {Aspelmeyer}},\ and\ \bibinfo
		{author} {\bibfnamefont {O.}~\bibnamefont {Painter}},\ }\bibfield  {title}
	{\bibinfo {title} {Squeezed light from a silicon micromechanical resonator},\
	}\href {https://doi.org/https://doi.org/10.1038/nature12307} {\bibfield
		{journal} {\bibinfo  {journal} {Nature}\ }\textbf {\bibinfo {volume} {500}},\
		\bibinfo {pages} {185} (\bibinfo {year} {2013})}\BibitemShut {NoStop}%
	\bibitem [{\citenamefont {Aggarwal}\ \emph {et~al.}(2020)\citenamefont
		{Aggarwal}, \citenamefont {Cullen}, \citenamefont {Cripe}, \citenamefont
		{Cole}, \citenamefont {Lanza}, \citenamefont {Libson}, \citenamefont
		{Follman}, \citenamefont {Heu}, \citenamefont {Corbitt},\ and\ \citenamefont
		{Mavalvala}}]{aggarwal2020room}%
	\BibitemOpen
	\bibfield  {author} {\bibinfo {author} {\bibfnamefont {N.}~\bibnamefont
			{Aggarwal}}, \bibinfo {author} {\bibfnamefont {T.~J.}\ \bibnamefont
			{Cullen}}, \bibinfo {author} {\bibfnamefont {J.}~\bibnamefont {Cripe}},
		\bibinfo {author} {\bibfnamefont {G.~D.}\ \bibnamefont {Cole}}, \bibinfo
		{author} {\bibfnamefont {R.}~\bibnamefont {Lanza}}, \bibinfo {author}
		{\bibfnamefont {A.}~\bibnamefont {Libson}}, \bibinfo {author} {\bibfnamefont
			{D.}~\bibnamefont {Follman}}, \bibinfo {author} {\bibfnamefont
			{P.}~\bibnamefont {Heu}}, \bibinfo {author} {\bibfnamefont {T.}~\bibnamefont
			{Corbitt}},\ and\ \bibinfo {author} {\bibfnamefont {N.}~\bibnamefont
			{Mavalvala}},\ }\bibfield  {title} {\bibinfo {title} {Room-temperature
			optomechanical squeezing},\ }\href
	{https://doi.org/https://doi.org/10.1038/s41567-020-0877-x} {\bibfield
		{journal} {\bibinfo  {journal} {Nat. Phys.}\ }\textbf {\bibinfo {volume}
			{16}},\ \bibinfo {pages} {784} (\bibinfo {year} {2020})}\BibitemShut
	{NoStop}%
	\bibitem [{\citenamefont {Li}\ \emph {et~al.}(2016)\citenamefont {Li},
		\citenamefont {Xuereb}, \citenamefont {Malossi},\ and\ \citenamefont
		{Vitali}}]{Li2016}%
	\BibitemOpen
	\bibfield  {author} {\bibinfo {author} {\bibfnamefont {J.}~\bibnamefont
			{Li}}, \bibinfo {author} {\bibfnamefont {A.}~\bibnamefont {Xuereb}}, \bibinfo
		{author} {\bibfnamefont {N.}~\bibnamefont {Malossi}},\ and\ \bibinfo {author}
		{\bibfnamefont {D.}~\bibnamefont {Vitali}},\ }\bibfield  {title} {\bibinfo
		{title} {Cavity mode frequencies and strong optomechanical coupling in
			two-membrane cavity optomechanics},\ }\href
	{https://doi.org/10.1088/2040-8978/18/8/084001} {\bibfield  {journal}
		{\bibinfo  {journal} {J. Opt.}\ }\textbf {\bibinfo {volume} {18}},\ \bibinfo
		{pages} {084001} (\bibinfo {year} {2016})}\BibitemShut {NoStop}%
	\bibitem [{\citenamefont {Newsom}\ \emph {et~al.}(2020)\citenamefont {Newsom},
		\citenamefont {Luna}, \citenamefont {Fedoseev}, \citenamefont {L{ö}ffler},\
		and\ \citenamefont {Bouwmeester}}]{Newsom2020}%
	\BibitemOpen
	\bibfield  {author} {\bibinfo {author} {\bibfnamefont {D.~C.}\ \bibnamefont
			{Newsom}}, \bibinfo {author} {\bibfnamefont {F.}~\bibnamefont {Luna}},
		\bibinfo {author} {\bibfnamefont {V.}~\bibnamefont {Fedoseev}}, \bibinfo
		{author} {\bibfnamefont {W.}~\bibnamefont {L{ö}ffler}},\ and\ \bibinfo
		{author} {\bibfnamefont {D.}~\bibnamefont {Bouwmeester}},\ }\bibfield
	{title} {\bibinfo {title} {Optimal optomechanical coupling strength in
			multi-membrane systems},\ }\href
	{https://doi.org/10.1103/PhysRevA.101.033829} {\bibfield  {journal} {\bibinfo
			{journal} {Phys. Rev. A}\ }\textbf {\bibinfo {volume} {101}},\ \bibinfo
		{pages} {033829} (\bibinfo {year} {2020})}\BibitemShut {NoStop}%
	\bibitem [{\citenamefont {Pfeifer}\ \emph {et~al.}(2022)\citenamefont
		{Pfeifer}, \citenamefont {Ratschbacher}, \citenamefont {Gallego},
		\citenamefont {Saavedra}, \citenamefont {Fa{\ss}bender}, \citenamefont {von
			Haaren}, \citenamefont {Alt}, \citenamefont {Hofferberth}, \citenamefont
		{K{\"o}hl}, \citenamefont {Linden} \emph {et~al.}}]{pfeifer2022achievements}%
	\BibitemOpen
	\bibfield  {author} {\bibinfo {author} {\bibfnamefont {H.}~\bibnamefont
			{Pfeifer}}, \bibinfo {author} {\bibfnamefont {L.}~\bibnamefont
			{Ratschbacher}}, \bibinfo {author} {\bibfnamefont {J.}~\bibnamefont
			{Gallego}}, \bibinfo {author} {\bibfnamefont {C.}~\bibnamefont {Saavedra}},
		\bibinfo {author} {\bibfnamefont {A.}~\bibnamefont {Fa{\ss}bender}}, \bibinfo
		{author} {\bibfnamefont {A.}~\bibnamefont {von Haaren}}, \bibinfo {author}
		{\bibfnamefont {W.}~\bibnamefont {Alt}}, \bibinfo {author} {\bibfnamefont
			{S.}~\bibnamefont {Hofferberth}}, \bibinfo {author} {\bibfnamefont
			{M.}~\bibnamefont {K{\"o}hl}}, \bibinfo {author} {\bibfnamefont
			{S.}~\bibnamefont {Linden}}, \emph {et~al.},\ }\bibfield  {title} {\bibinfo
		{title} {Achievements and perspectives of optical fiber fabry--perot
			cavities},\ }\href
	{https://doi.org/https://doi.org/10.1007/s00340-022-07752-8} {\bibfield
		{journal} {\bibinfo  {journal} {Applied Physics B}\ }\textbf {\bibinfo
			{volume} {128}},\ \bibinfo {pages} {29} (\bibinfo {year} {2022})}\BibitemShut
	{NoStop}%
	\bibitem [{\citenamefont {Saarinen}\ \emph {et~al.}(2023)\citenamefont
		{Saarinen}, \citenamefont {Kralj}, \citenamefont {Langman}, \citenamefont
		{Tsaturyan},\ and\ \citenamefont {Schliesser}}]{Saarinen:23}%
	\BibitemOpen
	\bibfield  {author} {\bibinfo {author} {\bibfnamefont {S.~A.}\ \bibnamefont
			{Saarinen}}, \bibinfo {author} {\bibfnamefont {N.}~\bibnamefont {Kralj}},
		\bibinfo {author} {\bibfnamefont {E.~C.}\ \bibnamefont {Langman}}, \bibinfo
		{author} {\bibfnamefont {Y.}~\bibnamefont {Tsaturyan}},\ and\ \bibinfo
		{author} {\bibfnamefont {A.}~\bibnamefont {Schliesser}},\ }\bibfield  {title}
	{\bibinfo {title} {Laser cooling a membrane-in-the-middle system close to the
			quantum ground state from room temperature},\ }\href
	{https://doi.org/10.1364/OPTICA.468590} {\bibfield  {journal} {\bibinfo
			{journal} {Optica}\ }\textbf {\bibinfo {volume} {10}},\ \bibinfo {pages}
		{364} (\bibinfo {year} {2023})}\BibitemShut {NoStop}%
	\bibitem [{\citenamefont {Norte}\ \emph {et~al.}(2016)\citenamefont {Norte},
		\citenamefont {Moura},\ and\ \citenamefont
		{Gr{\"o}blacher}}]{norte2016mechanical}%
	\BibitemOpen
	\bibfield  {author} {\bibinfo {author} {\bibfnamefont {R.~A.}\ \bibnamefont
			{Norte}}, \bibinfo {author} {\bibfnamefont {J.~P.}\ \bibnamefont {Moura}},\
		and\ \bibinfo {author} {\bibfnamefont {S.}~\bibnamefont {Gr{\"o}blacher}},\
	}\bibfield  {title} {\bibinfo {title} {Mechanical resonators for quantum
			optomechanics experiments at room temperature},\ }\href
	{https://doi.org/10.1103/PhysRevLett.116.147202} {\bibfield  {journal}
		{\bibinfo  {journal} {Phys. Rev. Lett.}\ }\textbf {\bibinfo {volume} {116}},\
		\bibinfo {pages} {147202} (\bibinfo {year} {2016})}\BibitemShut {NoStop}%
	\bibitem [{\citenamefont {Beccari}\ \emph {et~al.}(2022)\citenamefont
		{Beccari}, \citenamefont {Visani}, \citenamefont {Fedorov}, \citenamefont
		{Bereyhi}, \citenamefont {Boureau}, \citenamefont {Engelsen},\ and\
		\citenamefont {Kippenberg}}]{beccari2022strained}%
	\BibitemOpen
	\bibfield  {author} {\bibinfo {author} {\bibfnamefont {A.}~\bibnamefont
			{Beccari}}, \bibinfo {author} {\bibfnamefont {D.~A.}\ \bibnamefont {Visani}},
		\bibinfo {author} {\bibfnamefont {S.~A.}\ \bibnamefont {Fedorov}}, \bibinfo
		{author} {\bibfnamefont {M.~J.}\ \bibnamefont {Bereyhi}}, \bibinfo {author}
		{\bibfnamefont {V.}~\bibnamefont {Boureau}}, \bibinfo {author} {\bibfnamefont
			{N.~J.}\ \bibnamefont {Engelsen}},\ and\ \bibinfo {author} {\bibfnamefont
			{T.~J.}\ \bibnamefont {Kippenberg}},\ }\bibfield  {title} {\bibinfo {title}
		{Strained crystalline nanomechanical resonators with quality factors above 10
			billion},\ }\href
	{https://doi.org/https://doi.org/10.1038/s41567-021-01498-4} {\bibfield
		{journal} {\bibinfo  {journal} {Nat. Phys.}\ }\textbf {\bibinfo {volume}
			{18}},\ \bibinfo {pages} {436} (\bibinfo {year} {2022})}\BibitemShut
	{NoStop}%
	\bibitem [{\citenamefont {Tsaturyan}\ \emph {et~al.}(2017)\citenamefont
		{Tsaturyan}, \citenamefont {Barg}, \citenamefont {Polzik},\ and\
		\citenamefont {Schliesser}}]{tsaturyan2017ultracoherent}%
	\BibitemOpen
	\bibfield  {author} {\bibinfo {author} {\bibfnamefont {Y.}~\bibnamefont
			{Tsaturyan}}, \bibinfo {author} {\bibfnamefont {A.}~\bibnamefont {Barg}},
		\bibinfo {author} {\bibfnamefont {E.~S.}\ \bibnamefont {Polzik}},\ and\
		\bibinfo {author} {\bibfnamefont {A.}~\bibnamefont {Schliesser}},\ }\bibfield
	{title} {\bibinfo {title} {Ultracoherent nanomechanical resonators via soft
			clamping and dissipation dilution},\ }\href
	{https://doi.org/https://doi.org/10.1038/nnano.2017.101} {\bibfield
		{journal} {\bibinfo  {journal} {Nat. Nanotechnol.}\ }\textbf {\bibinfo
			{volume} {12}},\ \bibinfo {pages} {776} (\bibinfo {year} {2017})}\BibitemShut
	{NoStop}%
	\bibitem [{\citenamefont {Enzian}\ \emph {et~al.}(2023)\citenamefont {Enzian},
		\citenamefont {Wang}, \citenamefont {Simonsen}, \citenamefont {Mathiassen},
		\citenamefont {Vibel}, \citenamefont {Tsaturyan}, \citenamefont {Tagantsev},
		\citenamefont {Schliesser},\ and\ \citenamefont
		{Polzik}}]{enzian2023phononically}%
	\BibitemOpen
	\bibfield  {author} {\bibinfo {author} {\bibfnamefont {G.}~\bibnamefont
			{Enzian}}, \bibinfo {author} {\bibfnamefont {Z.}~\bibnamefont {Wang}},
		\bibinfo {author} {\bibfnamefont {A.}~\bibnamefont {Simonsen}}, \bibinfo
		{author} {\bibfnamefont {J.}~\bibnamefont {Mathiassen}}, \bibinfo {author}
		{\bibfnamefont {T.}~\bibnamefont {Vibel}}, \bibinfo {author} {\bibfnamefont
			{Y.}~\bibnamefont {Tsaturyan}}, \bibinfo {author} {\bibfnamefont
			{A.}~\bibnamefont {Tagantsev}}, \bibinfo {author} {\bibfnamefont
			{A.}~\bibnamefont {Schliesser}},\ and\ \bibinfo {author} {\bibfnamefont
			{E.~S.}\ \bibnamefont {Polzik}},\ }\bibfield  {title} {\bibinfo {title}
		{Phononically shielded photonic-crystal mirror membranes for cavity quantum
			optomechanics},\ }\href {https://doi.org/https://doi.org/10.1364/OE.484369}
	{\bibfield  {journal} {\bibinfo  {journal} {Opt. Express}\ }\textbf {\bibinfo
			{volume} {31}},\ \bibinfo {pages} {13040} (\bibinfo {year}
		{2023})}\BibitemShut {NoStop}%
	\bibitem [{\citenamefont {Gärtner}\ \emph {et~al.}(2018)\citenamefont
		{Gärtner}, \citenamefont {Moura}, \citenamefont {Haaxman}, \citenamefont
		{Norte},\ and\ \citenamefont {Gröblacher}}]{Gaertner2018}%
	\BibitemOpen
	\bibfield  {author} {\bibinfo {author} {\bibfnamefont {C.}~\bibnamefont
			{Gärtner}}, \bibinfo {author} {\bibfnamefont {J.~P.}\ \bibnamefont {Moura}},
		\bibinfo {author} {\bibfnamefont {W.}~\bibnamefont {Haaxman}}, \bibinfo
		{author} {\bibfnamefont {R.~A.}\ \bibnamefont {Norte}},\ and\ \bibinfo
		{author} {\bibfnamefont {S.}~\bibnamefont {Gröblacher}},\ }\bibfield
	{title} {\bibinfo {title} {Integrated optomechanical arrays of two high
			reflectivity {SiN} membranes},\ }\href
	{https://doi.org/10.1021/acs.nanolett.8b03240} {\bibfield  {journal}
		{\bibinfo  {journal} {Nano Lett.}\ }\textbf {\bibinfo {volume} {18}},\
		\bibinfo {pages} {7171} (\bibinfo {year} {2018})}\BibitemShut {NoStop}%
	\bibitem [{\citenamefont {Piergentili}\ \emph {et~al.}(2018)\citenamefont
		{Piergentili}, \citenamefont {Catalini}, \citenamefont {Bawaj}, \citenamefont
		{Zippilli}, \citenamefont {Malossi}, \citenamefont {Natali}, \citenamefont
		{Vitali},\ and\ \citenamefont {Di~Giuseppe}}]{Piergentili2018}%
	\BibitemOpen
	\bibfield  {author} {\bibinfo {author} {\bibfnamefont {P.}~\bibnamefont
			{Piergentili}}, \bibinfo {author} {\bibfnamefont {L.}~\bibnamefont
			{Catalini}}, \bibinfo {author} {\bibfnamefont {M.}~\bibnamefont {Bawaj}},
		\bibinfo {author} {\bibfnamefont {S.}~\bibnamefont {Zippilli}}, \bibinfo
		{author} {\bibfnamefont {N.}~\bibnamefont {Malossi}}, \bibinfo {author}
		{\bibfnamefont {R.}~\bibnamefont {Natali}}, \bibinfo {author} {\bibfnamefont
			{D.}~\bibnamefont {Vitali}},\ and\ \bibinfo {author} {\bibfnamefont
			{G.}~\bibnamefont {Di~Giuseppe}},\ }\bibfield  {title} {\bibinfo {title}
		{Two-membrane cavity optomechanics},\ }\href
	{https://doi.org/10.1088/1367-2630/aad85f} {\bibfield  {journal} {\bibinfo
			{journal} {New J. Phys.}\ }\textbf {\bibinfo {volume} {20}},\ \bibinfo
		{pages} {083024} (\bibinfo {year} {2018})}\BibitemShut {NoStop}%
	\bibitem [{\citenamefont {Stambaugh}\ \emph {et~al.}(2015)\citenamefont
		{Stambaugh}, \citenamefont {Xu}, \citenamefont {Kemiktarak}, \citenamefont
		{Taylor},\ and\ \citenamefont {Lawall}}]{stambaugh2015membrane}%
	\BibitemOpen
	\bibfield  {author} {\bibinfo {author} {\bibfnamefont {C.}~\bibnamefont
			{Stambaugh}}, \bibinfo {author} {\bibfnamefont {H.}~\bibnamefont {Xu}},
		\bibinfo {author} {\bibfnamefont {U.}~\bibnamefont {Kemiktarak}}, \bibinfo
		{author} {\bibfnamefont {J.}~\bibnamefont {Taylor}},\ and\ \bibinfo {author}
		{\bibfnamefont {J.}~\bibnamefont {Lawall}},\ }\bibfield  {title} {\bibinfo
		{title} {From membrane-in-the-middle to mirror-in-the-middle with a
			high-reflectivity sub-wavelength grating},\ }\href
	{https://doi.org/10.1002/andp.201400142} {\bibfield  {journal} {\bibinfo
			{journal} {Ann. Phys.}\ }\textbf {\bibinfo {volume} {527}},\ \bibinfo {pages}
		{81} (\bibinfo {year} {2015})}\BibitemShut {NoStop}%
	\bibitem [{\citenamefont {Riedinger}\ \emph {et~al.}(2018)\citenamefont
		{Riedinger}, \citenamefont {Wallucks}, \citenamefont {Marinković},
		\citenamefont {Löschnauer}, \citenamefont {Aspelmeyer}, \citenamefont
		{Hong},\ and\ \citenamefont {Gröblacher}}]{Riedinger2018}%
	\BibitemOpen
	\bibfield  {author} {\bibinfo {author} {\bibfnamefont {R.}~\bibnamefont
			{Riedinger}}, \bibinfo {author} {\bibfnamefont {A.}~\bibnamefont {Wallucks}},
		\bibinfo {author} {\bibfnamefont {I.}~\bibnamefont {Marinković}}, \bibinfo
		{author} {\bibfnamefont {C.}~\bibnamefont {Löschnauer}}, \bibinfo {author}
		{\bibfnamefont {M.}~\bibnamefont {Aspelmeyer}}, \bibinfo {author}
		{\bibfnamefont {S.}~\bibnamefont {Hong}},\ and\ \bibinfo {author}
		{\bibfnamefont {S.}~\bibnamefont {Gröblacher}},\ }\bibfield  {title}
	{\bibinfo {title} {Remote quantum entanglement between two micromechanical
			oscillators},\ }\href
	{https://doi.org/https://doi.org/10.1038/s41586-018-0036-z} {\bibfield
		{journal} {\bibinfo  {journal} {Nature}\ }\textbf {\bibinfo {volume} {556}},\
		\bibinfo {pages} {473} (\bibinfo {year} {2018})}\BibitemShut {NoStop}%
	\bibitem [{\citenamefont {Jayich}\ \emph {et~al.}(2008)\citenamefont {Jayich},
		\citenamefont {Sankey}, \citenamefont {Zwickl}, \citenamefont {Yang},
		\citenamefont {Thompson}, \citenamefont {Girvin}, \citenamefont {Clerk},
		\citenamefont {Marquardt},\ and\ \citenamefont
		{Harris}}]{jayich2008dispersive}%
	\BibitemOpen
	\bibfield  {author} {\bibinfo {author} {\bibfnamefont {A.}~\bibnamefont
			{Jayich}}, \bibinfo {author} {\bibfnamefont {J.}~\bibnamefont {Sankey}},
		\bibinfo {author} {\bibfnamefont {B.}~\bibnamefont {Zwickl}}, \bibinfo
		{author} {\bibfnamefont {C.}~\bibnamefont {Yang}}, \bibinfo {author}
		{\bibfnamefont {J.}~\bibnamefont {Thompson}}, \bibinfo {author}
		{\bibfnamefont {S.}~\bibnamefont {Girvin}}, \bibinfo {author} {\bibfnamefont
			{A.}~\bibnamefont {Clerk}}, \bibinfo {author} {\bibfnamefont
			{F.}~\bibnamefont {Marquardt}},\ and\ \bibinfo {author} {\bibfnamefont
			{J.}~\bibnamefont {Harris}},\ }\bibfield  {title} {\bibinfo {title}
		{Dispersive optomechanics: a membrane inside a cavity},\ }\href
	{https://doi.org/10.1088/1367-2630/10/9/095008} {\bibfield  {journal}
		{\bibinfo  {journal} {New J. Phys.}\ }\textbf {\bibinfo {volume} {10}},\
		\bibinfo {pages} {095008} (\bibinfo {year} {2008})}\BibitemShut {NoStop}%
	\bibitem [{\citenamefont {Wei}\ \emph {et~al.}(2019)\citenamefont {Wei},
		\citenamefont {Sheng}, \citenamefont {Yang}, \citenamefont {Wu},\ and\
		\citenamefont {Wu}}]{Wei2019PRA}%
	\BibitemOpen
	\bibfield  {author} {\bibinfo {author} {\bibfnamefont {X.}~\bibnamefont
			{Wei}}, \bibinfo {author} {\bibfnamefont {J.}~\bibnamefont {Sheng}}, \bibinfo
		{author} {\bibfnamefont {C.}~\bibnamefont {Yang}}, \bibinfo {author}
		{\bibfnamefont {Y.}~\bibnamefont {Wu}},\ and\ \bibinfo {author}
		{\bibfnamefont {H.}~\bibnamefont {Wu}},\ }\bibfield  {title} {\bibinfo
		{title} {Controllable two-membrane-in-the-middle cavity optomechanical
			system},\ }\href {https://doi.org/10.1103/PhysRevA.99.023851} {\bibfield
		{journal} {\bibinfo  {journal} {Phys. Rev. A}\ }\textbf {\bibinfo {volume}
			{99}},\ \bibinfo {pages} {023851} (\bibinfo {year} {2019})}\BibitemShut
	{NoStop}%
	\bibitem [{foo({\natexlab{a}})}]{footnote1}%
	\BibitemOpen
	\href@noop {} {\bibinfo {title} {Extracting this is equivalent to extracting
			${G} = \mathrm{max}(|\partial \omega_\mathrm{c} /\partial x|)$ for each
			dispersion curve, separated by the main cavity {FSR}.}}
	({\natexlab{a}})\BibitemShut {NoStop}%
	\bibitem [{\citenamefont {Nair}\ \emph {et~al.}(2017)\citenamefont {Nair},
		\citenamefont {Naesby},\ and\ \citenamefont {Dantan}}]{Nair2017}%
	\BibitemOpen
	\bibfield  {author} {\bibinfo {author} {\bibfnamefont {B.}~\bibnamefont
			{Nair}}, \bibinfo {author} {\bibfnamefont {A.}~\bibnamefont {Naesby}},\ and\
		\bibinfo {author} {\bibfnamefont {A.}~\bibnamefont {Dantan}},\ }\bibfield
	{title} {\bibinfo {title} {Optomechanical characterization of silicon nitride
			membrane arrays},\ }\href
	{https://doi.org/https://doi.org/10.1364/OL.42.001341} {\bibfield  {journal}
		{\bibinfo  {journal} {Opt. Lett.}\ }\textbf {\bibinfo {volume} {42}},\
		\bibinfo {pages} {1341} (\bibinfo {year} {2017})}\BibitemShut {NoStop}%
	\bibitem [{\citenamefont {Monticone}\ and\ \citenamefont
		{Alù}(2017)}]{Monticone2017}%
	\BibitemOpen
	\bibfield  {author} {\bibinfo {author} {\bibfnamefont {F.}~\bibnamefont
			{Monticone}}\ and\ \bibinfo {author} {\bibfnamefont {A.}~\bibnamefont
			{Alù}},\ }\bibfield  {title} {\bibinfo {title} {Bound states within the
			radiation continuum in diffraction gratings and the role of leaky modes},\
	}\href {https://doi.org/10.1088/1367-2630/aa849f} {\bibfield  {journal}
		{\bibinfo  {journal} {New J. Phys.}\ }\textbf {\bibinfo {volume} {19}},\
		\bibinfo {pages} {093011} (\bibinfo {year} {2017})}\BibitemShut {NoStop}%
	\bibitem [{\citenamefont {P{\'e}ralle}\ \emph {et~al.}(2024)\citenamefont
		{P{\'e}ralle}, \citenamefont {Manjeshwar}, \citenamefont {Ciers},
		\citenamefont {Wieczorek},\ and\ \citenamefont
		{Tassin}}]{peralle2024quasibound}%
	\BibitemOpen
	\bibfield  {author} {\bibinfo {author} {\bibfnamefont {C.}~\bibnamefont
			{P{\'e}ralle}}, \bibinfo {author} {\bibfnamefont {S.~K.}\ \bibnamefont
			{Manjeshwar}}, \bibinfo {author} {\bibfnamefont {A.}~\bibnamefont {Ciers}},
		\bibinfo {author} {\bibfnamefont {W.}~\bibnamefont {Wieczorek}},\ and\
		\bibinfo {author} {\bibfnamefont {P.}~\bibnamefont {Tassin}},\ }\bibfield
	{title} {\bibinfo {title} {Quasibound states in the continuum in photonic
			crystal based optomechanical microcavities},\ }\href
	{https://doi.org/10.1103/PhysRevB.109.035407} {\bibfield  {journal} {\bibinfo
			{journal} {Phys. Rev. B}\ }\textbf {\bibinfo {volume} {109}},\ \bibinfo
		{pages} {035407} (\bibinfo {year} {2024})}\BibitemShut {NoStop}%
	\bibitem [{\citenamefont {Lee}\ \emph {et~al.}(2002)\citenamefont {Lee},
		\citenamefont {Kim}, \citenamefont {Yoo}, \citenamefont {Hahn},\ and\
		\citenamefont {Lee}}]{lee2002spatial}%
	\BibitemOpen
	\bibfield  {author} {\bibinfo {author} {\bibfnamefont {J.~Y.}\ \bibnamefont
			{Lee}}, \bibinfo {author} {\bibfnamefont {J.~W.}\ \bibnamefont {Kim}},
		\bibinfo {author} {\bibfnamefont {Y.~S.}\ \bibnamefont {Yoo}}, \bibinfo
		{author} {\bibfnamefont {J.~W.}\ \bibnamefont {Hahn}},\ and\ \bibinfo
		{author} {\bibfnamefont {H.-W.}\ \bibnamefont {Lee}},\ }\bibfield  {title}
	{\bibinfo {title} {Spatial-domain cavity ringdown from a high-finesse plane
			fabry--perot cavity},\ }\href
	{https://doi.org/https://doi.org/10.1063/1.1425443} {\bibfield  {journal}
		{\bibinfo  {journal} {J. Appl. Phys.}\ }\textbf {\bibinfo {volume} {91}},\
		\bibinfo {pages} {582} (\bibinfo {year} {2002})}\BibitemShut {NoStop}%
	\bibitem [{\citenamefont {Sankey}\ \emph {et~al.}(2010)\citenamefont {Sankey},
		\citenamefont {Yang}, \citenamefont {Zwickl}, \citenamefont {Jayich},\ and\
		\citenamefont {Harris}}]{sankey2010strong}%
	\BibitemOpen
	\bibfield  {author} {\bibinfo {author} {\bibfnamefont {J.~C.}\ \bibnamefont
			{Sankey}}, \bibinfo {author} {\bibfnamefont {C.}~\bibnamefont {Yang}},
		\bibinfo {author} {\bibfnamefont {B.~M.}\ \bibnamefont {Zwickl}}, \bibinfo
		{author} {\bibfnamefont {A.~M.}\ \bibnamefont {Jayich}},\ and\ \bibinfo
		{author} {\bibfnamefont {J.~G.}\ \bibnamefont {Harris}},\ }\bibfield  {title}
	{\bibinfo {title} {Strong and tunable nonlinear optomechanical coupling in a
			low-loss system},\ }\href {https://doi.org/10.1038/nphys1707} {\bibfield
		{journal} {\bibinfo  {journal} {Nat. Phys.}\ }\textbf {\bibinfo {volume}
			{6}},\ \bibinfo {pages} {707} (\bibinfo {year} {2010})}\BibitemShut {NoStop}%
	\bibitem [{\citenamefont {Karuza}\ \emph {et~al.}(2012)\citenamefont {Karuza},
		\citenamefont {Galassi}, \citenamefont {Biancofiore}, \citenamefont
		{Molinelli}, \citenamefont {Natali}, \citenamefont {Tombesi}, \citenamefont
		{Di~Giuseppe},\ and\ \citenamefont {Vitali}}]{karuza2012tunable}%
	\BibitemOpen
	\bibfield  {author} {\bibinfo {author} {\bibfnamefont {M.}~\bibnamefont
			{Karuza}}, \bibinfo {author} {\bibfnamefont {M.}~\bibnamefont {Galassi}},
		\bibinfo {author} {\bibfnamefont {C.}~\bibnamefont {Biancofiore}}, \bibinfo
		{author} {\bibfnamefont {C.}~\bibnamefont {Molinelli}}, \bibinfo {author}
		{\bibfnamefont {R.}~\bibnamefont {Natali}}, \bibinfo {author} {\bibfnamefont
			{P.}~\bibnamefont {Tombesi}}, \bibinfo {author} {\bibfnamefont
			{G.}~\bibnamefont {Di~Giuseppe}},\ and\ \bibinfo {author} {\bibfnamefont
			{D.}~\bibnamefont {Vitali}},\ }\bibfield  {title} {\bibinfo {title} {Tunable
			linear and quadratic optomechanical coupling for a tilted membrane within an
			optical cavity: theory and experiment},\ }\href
	{https://doi.org/10.1088/2040-8978/15/2/025704}
	{\bibfield  {journal} {\bibinfo  {journal} {J. Opt.}\ }\textbf {\bibinfo
			{volume} {15}},\ \bibinfo {pages} {025704} (\bibinfo {year}
		{2012})}\BibitemShut {NoStop}%
	\bibitem [{\citenamefont {H\o{}j}\ \emph {et~al.}(2024)\citenamefont {H\o{}j},
		\citenamefont {Hoff},\ and\ \citenamefont {Andersen}}]{PhysRevX.14.011039}%
	\BibitemOpen
	\bibfield  {author} {\bibinfo {author} {\bibfnamefont {D.}~\bibnamefont
			{H\o{}j}}, \bibinfo {author} {\bibfnamefont {U.~B.}\ \bibnamefont {Hoff}},\
		and\ \bibinfo {author} {\bibfnamefont {U.~L.}\ \bibnamefont {Andersen}},\
	}\bibfield  {title} {\bibinfo {title} {Ultracoherent nanomechanical
			resonators based on density phononic crystal engineering},\ }\href
	{https://doi.org/10.1103/PhysRevX.14.011039} {\bibfield  {journal} {\bibinfo
			{journal} {Phys. Rev. X}\ }\textbf {\bibinfo {volume} {14}},\ \bibinfo
		{pages} {011039} (\bibinfo {year} {2024})}\BibitemShut {NoStop}%
	\bibitem [{\citenamefont {Planz}\ \emph {et~al.}(2023)\citenamefont {Planz},
		\citenamefont {Xi}, \citenamefont {Capelle}, \citenamefont {Langman},\ and\
		\citenamefont {Schliesser}}]{planz2023membrane}%
	\BibitemOpen
	\bibfield  {author} {\bibinfo {author} {\bibfnamefont {E.}~\bibnamefont
			{Planz}}, \bibinfo {author} {\bibfnamefont {X.}~\bibnamefont {Xi}}, \bibinfo
		{author} {\bibfnamefont {T.}~\bibnamefont {Capelle}}, \bibinfo {author}
		{\bibfnamefont {E.~C.}\ \bibnamefont {Langman}},\ and\ \bibinfo {author}
		{\bibfnamefont {A.}~\bibnamefont {Schliesser}},\ }\bibfield  {title}
	{\bibinfo {title} {Membrane-in-the-middle optomechanics with a soft-clamped
			membrane at millikelvin temperatures},\ }\href
	{https://doi.org/https://doi.org/10.1364/OE.502359} {\bibfield  {journal}
		{\bibinfo  {journal} {Opt. Express}\ }\textbf {\bibinfo {volume} {31}},\
		\bibinfo {pages} {41773} (\bibinfo {year} {2023})}\BibitemShut {NoStop}%
	\bibitem [{\citenamefont {Kitamura}\ \emph {et~al.}(2007)\citenamefont
		{Kitamura}, \citenamefont {Pilon},\ and\ \citenamefont
		{Jonasz}}]{kitamura2007optical}%
	\BibitemOpen
	\bibfield  {author} {\bibinfo {author} {\bibfnamefont {R.}~\bibnamefont
			{Kitamura}}, \bibinfo {author} {\bibfnamefont {L.}~\bibnamefont {Pilon}},\
		and\ \bibinfo {author} {\bibfnamefont {M.}~\bibnamefont {Jonasz}},\
	}\bibfield  {title} {\bibinfo {title} {Optical constants of silica glass from
			extreme ultraviolet to far infrared at near room temperature},\ }\href
	{https://doi.org/https://doi.org/10.1364/AO.46.008118} {\bibfield  {journal}
		{\bibinfo  {journal} {Appl. Opt.}\ }\textbf {\bibinfo {volume} {46}},\
		\bibinfo {pages} {8118} (\bibinfo {year} {2007})}\BibitemShut {NoStop}%
	\bibitem [{\citenamefont {McLemore}\ \emph {et~al.}(2022)\citenamefont
		{McLemore}, \citenamefont {Jin}, \citenamefont {Kelleher}, \citenamefont
		{Hendrie}, \citenamefont {Mason}, \citenamefont {Luo}, \citenamefont {Lee},
		\citenamefont {Rakich}, \citenamefont {Diddams},\ and\ \citenamefont
		{Quinlan}}]{PhysRevApplied.18.054054}%
	\BibitemOpen
	\bibfield  {author} {\bibinfo {author} {\bibfnamefont {C.~A.}\ \bibnamefont
			{McLemore}}, \bibinfo {author} {\bibfnamefont {N.}~\bibnamefont {Jin}},
		\bibinfo {author} {\bibfnamefont {M.~L.}\ \bibnamefont {Kelleher}}, \bibinfo
		{author} {\bibfnamefont {J.~P.}\ \bibnamefont {Hendrie}}, \bibinfo {author}
		{\bibfnamefont {D.}~\bibnamefont {Mason}}, \bibinfo {author} {\bibfnamefont
			{Y.}~\bibnamefont {Luo}}, \bibinfo {author} {\bibfnamefont {D.}~\bibnamefont
			{Lee}}, \bibinfo {author} {\bibfnamefont {P.}~\bibnamefont {Rakich}},
		\bibinfo {author} {\bibfnamefont {S.~A.}\ \bibnamefont {Diddams}},\ and\
		\bibinfo {author} {\bibfnamefont {F.}~\bibnamefont {Quinlan}},\ }\bibfield
	{title} {\bibinfo {title} {Miniaturizing ultrastable electromagnetic
			oscillators: Sub-${10}^{\ensuremath{-}14}$ frequency instability from a
			centimeter-scale fabry-perot cavity},\ }\href
	{https://doi.org/10.1103/PhysRevApplied.18.054054} {\bibfield  {journal}
		{\bibinfo  {journal} {Phys. Rev. Appl.}\ }\textbf {\bibinfo {volume} {18}},\
		\bibinfo {pages} {054054} (\bibinfo {year} {2022})}\BibitemShut {NoStop}%
	\bibitem [{\citenamefont {Kelleher}\ \emph {et~al.}(2023)\citenamefont
		{Kelleher}, \citenamefont {McLemore}, \citenamefont {Lee}, \citenamefont
		{Davila-Rodriguez}, \citenamefont {Diddams},\ and\ \citenamefont
		{Quinlan}}]{Kelleher:23}%
	\BibitemOpen
	\bibfield  {author} {\bibinfo {author} {\bibfnamefont {M.~L.}\ \bibnamefont
			{Kelleher}}, \bibinfo {author} {\bibfnamefont {C.~A.}\ \bibnamefont
			{McLemore}}, \bibinfo {author} {\bibfnamefont {D.}~\bibnamefont {Lee}},
		\bibinfo {author} {\bibfnamefont {J.}~\bibnamefont {Davila-Rodriguez}},
		\bibinfo {author} {\bibfnamefont {S.~A.}\ \bibnamefont {Diddams}},\ and\
		\bibinfo {author} {\bibfnamefont {F.}~\bibnamefont {Quinlan}},\ }\bibfield
	{title} {\bibinfo {title} {Compact, portable, thermal-noise-limited optical
			cavity with low acceleration sensitivity},\ }\href
	{https://doi.org/10.1364/OE.486087} {\bibfield  {journal} {\bibinfo
			{journal} {Opt. Express}\ }\textbf {\bibinfo {volume} {31}},\ \bibinfo
		{pages} {11954} (\bibinfo {year} {2023})}\BibitemShut {NoStop}%
	\bibitem [{\citenamefont {Rossi}\ \emph {et~al.}(2018)\citenamefont {Rossi},
		\citenamefont {Mason}, \citenamefont {Chen}, \citenamefont {Tsaturyan},\ and\
		\citenamefont {Schliesser}}]{rossi2018measurement}%
	\BibitemOpen
	\bibfield  {author} {\bibinfo {author} {\bibfnamefont {M.}~\bibnamefont
			{Rossi}}, \bibinfo {author} {\bibfnamefont {D.}~\bibnamefont {Mason}},
		\bibinfo {author} {\bibfnamefont {J.}~\bibnamefont {Chen}}, \bibinfo {author}
		{\bibfnamefont {Y.}~\bibnamefont {Tsaturyan}},\ and\ \bibinfo {author}
		{\bibfnamefont {A.}~\bibnamefont {Schliesser}},\ }\bibfield  {title}
	{\bibinfo {title} {Measurement-based quantum control of mechanical motion},\
	}\href {https://doi.org/https://doi.org/10.1038/s41586-018-0643-8} {\bibfield
		{journal} {\bibinfo  {journal} {Nature}\ }\textbf {\bibinfo {volume}
			{563}},\ \bibinfo {pages} {53} (\bibinfo {year} {2018})}\BibitemShut
	{NoStop}%
	\bibitem [{\citenamefont {Dumont}\ \emph {et~al.}(2019)\citenamefont {Dumont},
		\citenamefont {Bernard}, \citenamefont {Reinhardt}, \citenamefont {Kato},
		\citenamefont {Ruf},\ and\ \citenamefont {Sankey}}]{dumont2019flexure}%
	\BibitemOpen
	\bibfield  {author} {\bibinfo {author} {\bibfnamefont {V.}~\bibnamefont
			{Dumont}}, \bibinfo {author} {\bibfnamefont {S.}~\bibnamefont {Bernard}},
		\bibinfo {author} {\bibfnamefont {C.}~\bibnamefont {Reinhardt}}, \bibinfo
		{author} {\bibfnamefont {A.}~\bibnamefont {Kato}}, \bibinfo {author}
		{\bibfnamefont {M.}~\bibnamefont {Ruf}},\ and\ \bibinfo {author}
		{\bibfnamefont {J.~C.}\ \bibnamefont {Sankey}},\ }\bibfield  {title}
	{\bibinfo {title} {Flexure-tuned membrane-at-the-edge optomechanical
			system},\ }\href {https://doi.org/https://doi.org/10.1364/OE.27.025731}
	{\bibfield  {journal} {\bibinfo  {journal} {Opt. Express}\ }\textbf {\bibinfo
			{volume} {27}},\ \bibinfo {pages} {25731} (\bibinfo {year}
		{2019})}\BibitemShut {NoStop}%
	\bibitem [{\citenamefont {Moura}\ \emph {et~al.}(2018)\citenamefont {Moura},
		\citenamefont {Norte}, \citenamefont {Guo}, \citenamefont
		{Sch{\"a}fermeier},\ and\ \citenamefont
		{Gr{\"o}blacher}}]{moura2018centimeter}%
	\BibitemOpen
	\bibfield  {author} {\bibinfo {author} {\bibfnamefont {J.~P.}\ \bibnamefont
			{Moura}}, \bibinfo {author} {\bibfnamefont {R.~A.}\ \bibnamefont {Norte}},
		\bibinfo {author} {\bibfnamefont {J.}~\bibnamefont {Guo}}, \bibinfo {author}
		{\bibfnamefont {C.}~\bibnamefont {Sch{\"a}fermeier}},\ and\ \bibinfo {author}
		{\bibfnamefont {S.}~\bibnamefont {Gr{\"o}blacher}},\ }\bibfield  {title}
	{\bibinfo {title} {Centimeter-scale suspended photonic crystal mirrors},\
	}\href {https://doi.org/10.1364/OE.26.001895} {\bibfield  {journal} {\bibinfo
			{journal} {Opt. Express}\ }\textbf {\bibinfo {volume} {26}},\ \bibinfo
		{pages} {1895} (\bibinfo {year} {2018})}\BibitemShut {NoStop}%
	\bibitem [{\citenamefont {Bao}\ \emph {et~al.}(2002)\citenamefont {Bao},
		\citenamefont {Yang}, \citenamefont {Yin},\ and\ \citenamefont
		{Sun}}]{bao2002energy}%
	\BibitemOpen
	\bibfield  {author} {\bibinfo {author} {\bibfnamefont {M.}~\bibnamefont
			{Bao}}, \bibinfo {author} {\bibfnamefont {H.}~\bibnamefont {Yang}}, \bibinfo
		{author} {\bibfnamefont {H.}~\bibnamefont {Yin}},\ and\ \bibinfo {author}
		{\bibfnamefont {Y.}~\bibnamefont {Sun}},\ }\bibfield  {title} {\bibinfo
		{title} {Energy transfer model for squeeze-film air damping in low vacuum},\
	}\href {https://doi.org/10.1088/0960-1317/12/3/322} {\bibfield  {journal}
		{\bibinfo  {journal} {J. Micromech. Microeng.}\ }\textbf {\bibinfo {volume}
			{12}},\ \bibinfo {pages} {341} (\bibinfo {year} {2002})}\BibitemShut
	{NoStop}%
	\bibitem [{\citenamefont {Stanley}\ \emph {et~al.}(1994)\citenamefont
		{Stanley}, \citenamefont {Houdre}, \citenamefont {Oesterle}, \citenamefont
		{Gailhanou},\ and\ \citenamefont {Ilegems}}]{stanley1994ultrahigh}%
	\BibitemOpen
	\bibfield  {author} {\bibinfo {author} {\bibfnamefont {R.}~\bibnamefont
			{Stanley}}, \bibinfo {author} {\bibfnamefont {R.}~\bibnamefont {Houdre}},
		\bibinfo {author} {\bibfnamefont {U.}~\bibnamefont {Oesterle}}, \bibinfo
		{author} {\bibfnamefont {M.}~\bibnamefont {Gailhanou}},\ and\ \bibinfo
		{author} {\bibfnamefont {M.}~\bibnamefont {Ilegems}},\ }\bibfield  {title}
	{\bibinfo {title} {Ultrahigh finesse microcavity with distributed bragg
			reflectors},\ }\href {https://doi.org/10.1063/1.112877} {\bibfield  {journal}
		{\bibinfo  {journal} {Appl. Phys. Lett.}\ }\textbf {\bibinfo {volume} {65}},\
		\bibinfo {pages} {1883} (\bibinfo {year} {1994})}\BibitemShut {NoStop}%
	\bibitem [{\citenamefont {Black}(2001)}]{Black2001}%
	\BibitemOpen
	\bibfield  {author} {\bibinfo {author} {\bibfnamefont {E.~D.}\ \bibnamefont
			{Black}},\ }\bibfield  {title} {\bibinfo {title} {An introduction to
			{Pound-Drever-Hall} laser frequency stabilization},\ }\href
	{https://doi.org/10.1119/1.1286663} {\bibfield  {journal} {\bibinfo
			{journal} {Am.\ J.\ Phys.}\ }\textbf {\bibinfo {volume} {69}},\ \bibinfo
		{pages} {79} (\bibinfo {year} {2001})}\BibitemShut {NoStop}%
	\bibitem [{\citenamefont {Fan}\ and\ \citenamefont
		{Joannopoulos}(2002)}]{fan2002analysis}%
	\BibitemOpen
	\bibfield  {author} {\bibinfo {author} {\bibfnamefont {S.}~\bibnamefont
			{Fan}}\ and\ \bibinfo {author} {\bibfnamefont {J.~D.}\ \bibnamefont
			{Joannopoulos}},\ }\bibfield  {title} {\bibinfo {title} {Analysis of guided
			resonances in photonic crystal slabs},\ }\href
	{https://doi.org/10.1103/PhysRevB.65.235112} {\bibfield  {journal} {\bibinfo
			{journal} {Phys. Rev. B}\ }\textbf {\bibinfo {volume} {65}},\ \bibinfo
		{pages} {235112} (\bibinfo {year} {2002})}\BibitemShut {NoStop}%
	\bibitem [{\citenamefont {Cheung}\ and\ \citenamefont
		{Law}(2011)}]{Cheung2011MIM}%
	\BibitemOpen
	\bibfield  {author} {\bibinfo {author} {\bibfnamefont {H.~K.}\ \bibnamefont
			{Cheung}}\ and\ \bibinfo {author} {\bibfnamefont {C.~K.}\ \bibnamefont
			{Law}},\ }\bibfield  {title} {\bibinfo {title} {Nonadiabatic optomechanical
			hamiltonian of a moving dielectric membrane in a cavity},\ }\href
	{https://doi.org/10.1103/PhysRevA.84.023812} {\bibfield  {journal} {\bibinfo
			{journal} {Phys. Rev. A}\ }\textbf {\bibinfo {volume} {84}},\ \bibinfo
		{pages} {023812} (\bibinfo {year} {2011})}\BibitemShut {NoStop}%
	\bibitem [{\citenamefont {Pedrotti}\ \emph {et~al.}(2017)\citenamefont
		{Pedrotti}, \citenamefont {Pedrotti},\ and\ \citenamefont
		{Pedrotti}}]{pedrotti2017introduction}%
	\BibitemOpen
	\bibfield  {author} {\bibinfo {author} {\bibfnamefont {F.~L.}\ \bibnamefont
			{Pedrotti}}, \bibinfo {author} {\bibfnamefont {L.~M.}\ \bibnamefont
			{Pedrotti}},\ and\ \bibinfo {author} {\bibfnamefont {L.~S.}\ \bibnamefont
			{Pedrotti}},\ }\href@noop {} {\emph {\bibinfo {title} {Introduction to
				optics}}}\ (\bibinfo  {publisher} {Cambridge University Press},\ \bibinfo
	{year} {2017})\BibitemShut {NoStop}%
	\bibitem [{\citenamefont {Brooker}(2003)}]{brooker2003modern}%
	\BibitemOpen
	\bibfield  {author} {\bibinfo {author} {\bibfnamefont {G.}~\bibnamefont
			{Brooker}},\ }\href@noop {} {\emph {\bibinfo {title} {Modern classical
				optics}}},\ Vol.~\bibinfo {volume} {8}\ (\bibinfo  {publisher} {Oxford
		University Press},\ \bibinfo {year} {2003})\BibitemShut {NoStop}%
	\bibitem [{\citenamefont {Sheng}\ \emph
		{et~al.}(2020{\natexlab{b}})\citenamefont {Sheng}, \citenamefont {Wei},
		\citenamefont {Yang},\ and\ \citenamefont {Wu}}]{sheng2020self}%
	\BibitemOpen
	\bibfield  {author} {\bibinfo {author} {\bibfnamefont {J.}~\bibnamefont
			{Sheng}}, \bibinfo {author} {\bibfnamefont {X.}~\bibnamefont {Wei}}, \bibinfo
		{author} {\bibfnamefont {C.}~\bibnamefont {Yang}},\ and\ \bibinfo {author}
		{\bibfnamefont {H.}~\bibnamefont {Wu}},\ }\bibfield  {title} {\bibinfo
		{title} {Self-organized synchronization of phonon lasers},\ }\href
	{https://doi.org/10.1103/PhysRevLett.124.053604} {\bibfield  {journal}
		{\bibinfo  {journal} {Phys. Rev. Lett.}\ }\textbf {\bibinfo {volume} {124}},\
		\bibinfo {pages} {053604} (\bibinfo {year} {2020}{\natexlab{b}})}\BibitemShut
	{NoStop}%
	\bibitem [{\citenamefont {Poot}\ and\ \citenamefont {{van der
				Zant}}(2012)}]{POOT2012273}%
	\BibitemOpen
	\bibfield  {author} {\bibinfo {author} {\bibfnamefont {M.}~\bibnamefont
			{Poot}}\ and\ \bibinfo {author} {\bibfnamefont {H.~S.}\ \bibnamefont {{van
					der Zant}}},\ }\bibfield  {title} {\bibinfo {title} {Mechanical systems in
			the quantum regime},\ }\href {https://doi.org/10.1016/j.physrep.2011.12.004}
	{\bibfield  {journal} {\bibinfo  {journal} {Phys. Rep.}\ }\textbf {\bibinfo
			{volume} {511}},\ \bibinfo {pages} {273} (\bibinfo {year}
		{2012})}\BibitemShut {NoStop}%
	\bibitem [{\citenamefont {de~Jong}\ \emph {et~al.}(2023)\citenamefont
		{de~Jong}, \citenamefont {Cupertino}, \citenamefont {Shin}, \citenamefont
		{Gr\"oblacher}, \citenamefont {Alijani}, \citenamefont {Steeneken},\ and\
		\citenamefont {Norte}}]{Matthijspra2023}%
	\BibitemOpen
	\bibfield  {author} {\bibinfo {author} {\bibfnamefont {M.~H.}\ \bibnamefont
			{de~Jong}}, \bibinfo {author} {\bibfnamefont {A.}~\bibnamefont {Cupertino}},
		\bibinfo {author} {\bibfnamefont {D.}~\bibnamefont {Shin}}, \bibinfo {author}
		{\bibfnamefont {S.}~\bibnamefont {Gr\"oblacher}}, \bibinfo {author}
		{\bibfnamefont {F.}~\bibnamefont {Alijani}}, \bibinfo {author} {\bibfnamefont
			{P.~G.}\ \bibnamefont {Steeneken}},\ and\ \bibinfo {author} {\bibfnamefont
			{R.~A.}\ \bibnamefont {Norte}},\ }\bibfield  {title} {\bibinfo {title}
		{Beating ringdowns of near-degenerate mechanical resonances},\ }\href
	{https://doi.org/10.1103/PhysRevApplied.20.024053} {\bibfield  {journal}
		{\bibinfo  {journal} {Phys. Rev. Appl.}\ }\textbf {\bibinfo {volume} {20}},\
		\bibinfo {pages} {024053} (\bibinfo {year} {2023})}\BibitemShut {NoStop}%
	\bibitem [{foo({\natexlab{b}})}]{footnote2}%
	\BibitemOpen
	\href@noop {} {\bibinfo {title} {It is important to note that even for weakly
			coupled resonators, solving {E}qs.~\eqref{eq: TMM_eom} yields eigenmodes with
			frequencies that slightly differ from the membranes' intrinsic
			eigenfrequencies. {H}owever, because this difference is minimal in this
			regime, we maintain the same notation for the frequencies.}}
	({\natexlab{b}})\BibitemShut {NoStop}%
	\bibitem [{\citenamefont {Elste}\ \emph {et~al.}(2009)\citenamefont {Elste},
		\citenamefont {Girvin},\ and\ \citenamefont {Clerk}}]{elste2009quantum}%
	\BibitemOpen
	\bibfield  {author} {\bibinfo {author} {\bibfnamefont {F.}~\bibnamefont
			{Elste}}, \bibinfo {author} {\bibfnamefont {S.}~\bibnamefont {Girvin}},\ and\
		\bibinfo {author} {\bibfnamefont {A.}~\bibnamefont {Clerk}},\ }\bibfield
	{title} {\bibinfo {title} {Quantum noise interference and backaction cooling
			in cavity nanomechanics},\ }\href
	{https://doi.org/10.1103/PhysRevLett.102.207209} {\bibfield  {journal}
		{\bibinfo  {journal} {Phys. Rev. Lett.}\ }\textbf {\bibinfo {volume} {102}},\
		\bibinfo {pages} {207209} (\bibinfo {year} {2009})}\BibitemShut {NoStop}%
	\bibitem [{\citenamefont {Fitzgerald}\ \emph {et~al.}(2021)\citenamefont
		{Fitzgerald}, \citenamefont {Manjeshwar}, \citenamefont {Wieczorek},\ and\
		\citenamefont {Tassin}}]{fitzgerald2021cavity}%
	\BibitemOpen
	\bibfield  {author} {\bibinfo {author} {\bibfnamefont {J.~M.}\ \bibnamefont
			{Fitzgerald}}, \bibinfo {author} {\bibfnamefont {S.~K.}\ \bibnamefont
			{Manjeshwar}}, \bibinfo {author} {\bibfnamefont {W.}~\bibnamefont
			{Wieczorek}},\ and\ \bibinfo {author} {\bibfnamefont {P.}~\bibnamefont
			{Tassin}},\ }\bibfield  {title} {\bibinfo {title} {Cavity optomechanics with
			photonic bound states in the continuum},\ }\href
	{https://doi.org/10.1103/PhysRevResearch.3.013131} {\bibfield  {journal}
		{\bibinfo  {journal} {Phys. Rev. Research}\ }\textbf {\bibinfo {volume}
			{3}},\ \bibinfo {pages} {013131} (\bibinfo {year} {2021})}\BibitemShut
	{NoStop}%
	\bibitem [{\citenamefont {Monsel}\ \emph {et~al.}(2024)\citenamefont {Monsel},
		\citenamefont {Ciers}, \citenamefont {Manjeshwar}, \citenamefont
		{Wieczorek},\ and\ \citenamefont {Splettstoesser}}]{monsel2023dissipative}%
	\BibitemOpen
	\bibfield  {author} {\bibinfo {author} {\bibfnamefont {J.}~\bibnamefont
			{Monsel}}, \bibinfo {author} {\bibfnamefont {A.}~\bibnamefont {Ciers}},
		\bibinfo {author} {\bibfnamefont {S.~K.}\ \bibnamefont {Manjeshwar}},
		\bibinfo {author} {\bibfnamefont {W.}~\bibnamefont {Wieczorek}},\ and\
		\bibinfo {author} {\bibfnamefont {J.}~\bibnamefont {Splettstoesser}},\
	}\bibfield  {title} {\bibinfo {title} {Dissipative and dispersive cavity
			optomechanics with a frequency-dependent mirror},\ }\href
	{https://doi.org/10.1103/PhysRevA.109.043532} {\bibfield  {journal} {\bibinfo
			{journal} {Phys. Rev. A}\ }\textbf {\bibinfo {volume} {109}},\ \bibinfo
		{pages} {043532} (\bibinfo {year} {2024})}\BibitemShut {NoStop}%
	\bibitem [{\citenamefont {Guo}\ \emph {et~al.}(2017)\citenamefont {Guo},
		\citenamefont {Norte},\ and\ \citenamefont
		{Gr\"{o}blacher}}]{guo2017integrated}%
	\BibitemOpen
	\bibfield  {author} {\bibinfo {author} {\bibfnamefont {J.}~\bibnamefont
			{Guo}}, \bibinfo {author} {\bibfnamefont {R.~A.}\ \bibnamefont {Norte}},\
		and\ \bibinfo {author} {\bibfnamefont {S.}~\bibnamefont {Gr\"{o}blacher}},\
	}\bibfield  {title} {\bibinfo {title} {Integrated optical force sensors using
			focusing photonic crystal arrays},\ }\href
	{https://doi.org/10.1364/OE.25.009196} {\bibfield  {journal} {\bibinfo
			{journal} {Opt. Express}\ }\textbf {\bibinfo {volume} {25}},\ \bibinfo
		{pages} {9196} (\bibinfo {year} {2017})}\BibitemShut {NoStop}%
	\bibitem [{\citenamefont {Agrawal}\ \emph {et~al.}(2024)\citenamefont
		{Agrawal}, \citenamefont {Manley}, \citenamefont {Allepuz-Requena},\ and\
		\citenamefont {Wilson}}]{Agrawal:24}%
	\BibitemOpen
	\bibfield  {author} {\bibinfo {author} {\bibfnamefont {A.~R.}\ \bibnamefont
			{Agrawal}}, \bibinfo {author} {\bibfnamefont {J.}~\bibnamefont {Manley}},
		\bibinfo {author} {\bibfnamefont {D.}~\bibnamefont {Allepuz-Requena}},\ and\
		\bibinfo {author} {\bibfnamefont {D.~J.}\ \bibnamefont {Wilson}},\ }\bibfield
	{title} {\bibinfo {title} {Focusing membrane metamirrors for integrated
			cavity optomechanics},\ }\href {https://doi.org/10.1364/OPTICA.522509}
	{\bibfield  {journal} {\bibinfo  {journal} {Optica}\ }\textbf {\bibinfo
			{volume} {11}},\ \bibinfo {pages} {1235} (\bibinfo {year}
		{2024})}\BibitemShut {NoStop}%
\end{thebibliography}
\end{document}